\documentclass[11pt]{article}
\usepackage[paperwidth=8.5in,paperheight=11in,top=1in, bottom=1in, left=1in, right=1in]{geometry}

\usepackage[title]{appendix}
\usepackage[dvipsnames]{xcolor}
\usepackage{amsfonts,amsthm,amssymb}
\usepackage{dsfont}
\usepackage{tikz}
\usepackage{setspace}
\usepackage{subfigure}
\usepackage{booktabs} 
\usepackage{multirow}
\usepackage[flushleft]{threeparttable}
\usepackage{epstopdf}
\usepackage{setspace} 
\usepackage{floatrow}
\usepackage{mathtools}
\mathtoolsset{showonlyrefs=true}
\usepackage{float}
\usepackage{graphicx}
\usepackage{color}
\usepackage{indentfirst}
\usepackage{soul}

\usepackage[authoryear]{natbib}

\usepackage[hypertexnames=false, plainpages=false, pdfpagelabels]{hyperref} 

\newcommand\be{\begin{equation}}
	\newcommand\ee{\end{equation}}

\begin{document}
	\begin{center}

		\Large Net Buying Pressure and the Information in Bitcoin Option Trades\\ 
		
		\vspace{2ex}

		 		\normalsize Carol Alexander,$^{**}$  Jun Deng,$^*$ Jianfen Feng,$^*$ and Huning Wan$^*$\\ 
		
		
		This Version: \today

		\vspace{2ex}
		
		\begin{abstract}
			\noindent How do supply and demand from informed traders drive market prices of bitcoin options? Deribit options tick-level data supports the limits-to-arbitrage hypothesis about the market maker’s supply. The main demand-side effects are that at-the-money option prices are largely driven by volatility traders and out-of-the-money options are simultaneously driven by volatility traders and those with proprietary information about the direction of future bitcoin price movements.  The demand-side trading results contrast with prior studies on established options markets in the US and Asia, but we also show that Deribit is rapidly evolving into a more efficient channel for aggregating information from informed traders.
		\end{abstract}
	\end{center}
	\vspace{2ex}

	\vspace{2ex}

	 	\vspace{2ex}
	 	\singlespacing
	 	\noindent ** \small Corresponding author. University of Sussex Business School and Peking University HSBC Business School. Jubilee Building, University of Sussex, Falmer, Brighton, BN1 9RH, UK. \\ \url{c.alexander@sussex.ac.uk}\\ 
	
	 	\noindent * School of Banking and Finance, University of International Business and Economics. No.10, Huixin Dongjie, Chaoyang District, Beijing, 100029, China. \\\url{jundeng@uibe.edu.cn}; \\ \url{jianfen_feng@uibe.edu.cn}; \\\url{wanhuning@126.com}.
	
	\newpage
	\doublespacing
	\section{Introduction}
	\noindent {Bitcoin options markets are attracting a growing number of sophisticated and informed traders.   Well-known asset managers such as XBTO have been making markets in bitcoin options for many years, particularly on the} Deribit exchange, which is much the largest of the popular but self-regulated crypto options trading platforms {with over one billion USD daily trading volume}.\footnote{{See the websites of \href{https://www.xbto.com/}{XBTO} and \href{www.deribit.com}{Deribit}.}} Also, several large institutions have recently launched managed bitcoin funds that plan to charge high fees in return for active strategies such as market timing and strategic hedging using bitcoin options.\footnote{See, for example,  \href{https://www.coindesk.com/jpmorgan-to-let-clients-invest-in-bitcoin-fund-for-first-time-sources}{JP Morgan, CoinDesk, 26 April 2021.}}  It is therefore important for  regulators, and for the growing crypto investment community,  to understand the information drivers of bitcoin option prices.  Our study aims to fill this important gap in the literature.
	
	A primary role for financial markets is to aggregate and process information across different trading venues and to disperse it between market participants. {If bitcoin options market could aggregate and process volatility and directional information effectively, the regulators and traders would trust that the market attracts more sophisticated participants and  is functioning well.}
	The information is revealed through trades and prices that are shaped by the micro-structure of the market and   different types of professional traders, including liquidity providers, arbitragers and brokers. {The fundamental value (if any) and market micro-structure of the Blockchain based cryptocurrency are quite different from traditional stock and index markets.}  The classic work by \cite{easley1998option} shows  that two important determinants of informed trading are the market depth  and  the potential leverage available on the platform. Since then, the effect of such trading on prices and information flows has attracted wide-spread academic interest focusing on options markets, which allow for particularly fluid information channels because of the high leverage, low transaction costs and the absence of short‐selling restrictions.   { Bitcoin option market makers and traders would benefit from knowing the mainly driving force for different moneyness options that helps them to manage inventory and  place orders.}

	A widely-observed phenomenon in options markets is the smile or skew shape of a fixed-maturity implied volatility curve as a function of the moneyness (or strike) of the options. Following  pioneering work on US markets by  \cite{bollen2004does},  it is possible to infer which types of informed traders are dominating  in an options market by relating the observed movements in the implied volatility smile to a metric called \textit{net buying pressure}, defined as a delta-weighted difference between  buyer-initiated and seller-initiated option trades. If the options markets were  frictionless and complete, professional arbitrageurs could hedge positions perfectly and so the market maker's supply curves should be independent of the quantity traded. In this case movements in the implied volatility smile surface should be unrelated to the demand for options. However, in reality there are many restrictions that prevent perfect hedging. The costs of hedging increase with  the imbalance of  a market maker's net positions, and so does liquidity risk. To cover these costs and risks they are forced to charge a risk premium. In other words, the market maker's options supply curves slope upwards. The \textit{limits-to-arbitrage} hypothesis assumes the  market maker would manage inventory at a rational level, so changes in net demand would affect option prices and hence also their implied volatilities. In particular,  if at any given moneyness level the net buying pressure is positive (negative), the corresponding implied volatilities would rise (fall). Under this hypothesis, when changes in implied volatility are induced by excessive buying pressures in demand, these would reverse when the market maker  rebalances his portfolio. 
	
	\cite{bollen2004does} suggest an alternative \textit{learning hypothesis} whereby the market maker learns about underlying price dynamics from the trading activity of traders, and he uses this knowledge to update prices. This way,  implied volatility dynamics will respond to changes in demand  from informed option traders. In this first \textit{volatility learning} version of the learning hypothesis, traders are only assumed to trade on their knowledge about  volatility, not  the direction of change in the underlying asset price. This assumption rests on the observation that transaction costs are much higher in options  than they are in spot and futures markets, so directional information is easier to exploit by trading in spot and futures than it is by trading in options. However, volatility can only be traded in options markets, using strategies like butterfly spreads, straddles or strangles.
	
	Regression models devised by \cite{bollen2004does} test for both limits-to-arbitrage and volatility/directional learning  using regressions of implied volatility changes on different measures of net buying pressure and lagged changes in implied volatility, controlling for well-known features such as leverage effects that are independent of either hypothesis. Changes in implied volatility should be negatively autocorrelated under the limits-to-arbitrage hypothesis but have no significant autocorrelation under the volatility-learning hypothesis. A second regression-based test to distinguish the two hypotheses looks specifically at the demand for call and put options that are close to at-the-money (ATM). These options have the greatest demand from informed volatility traders because they have the highest volatility sensitivity. Under the volatility-learning hypothesis, where the market maker learns most from ATM options demand, the rest of the smile should move in tandem with ATM implied volatilities. For instance, if volatility is anticipated to rise (fall) then volatility-informed traders would buy (sell) straddles, so the demand for calls should behave similarly to the demand for puts. This way, the entire smile should move along with ATM volatility. By contrast, under the limits-to-arbitrage hypothesis the implied volatility in each moneyness category need not be related to changes in ATM volatility.

	Following \cite{kang2008information} a later strand of the literature hypothesizes the co-existence of two different types of informed traders. In addition to the \textit{volatility-motivated} type, who has advanced information about volatility,  the \textit{directional-motivated} type possess privileged information about the direction of future movements in the price of the underlying asset.  Therefore, if we  assume that traders in options markets have the most sophisticated access to privileged information, any finding of volatility and/or directional motivated buying pressure could be used by market makers to infer likely movements in future (such as a rise in price or a reduction in volatility) as a signal to rebalance their portfolios. The two distinct types of informed traders have strong incentives to trade options, but with different order imbalances, and therefore also different types of buying pressure on the market maker, and consequently different impacts on implied volatility changes. Again, and by contrast with the limits-to-arbitrage hypothesis,  both these hypotheses predict no autocorrelation in the dynamics of implied volatility surfaces.

	\cite{kang2008information} and \cite{chen2017net} extend the regression specification to include other buying pressure metrics and examine how ATM options demand affects the demand for options in other moneyness categories. When the informed traders are volatility-motivated, but have no knowledge about the direction of price changes, volatility shocks would shift the options supply curves and consequently also affect the smile. In this case, and as already argued above, it is the demand for ATM options that should mainly drive the entire smile and the implied volatilities of other options should move alongside. 
	By contrast, when the informed traders are directional-motivated but have no knowledge about the size of future price changes, this generates different buying pressures on calls relative to puts. For instance, if the price is thought to rise, directional-informed traders would buy calls and sell puts so the net buying pressure would be  positive on calls and negative  on puts.  

	The extant empirical literature only examines established options markets. Some papers confirm the limits-to-arbitrage hypothesis  and find no evidence to support either learning hypothesis. These include:  \cite{bollen2004does} for S\&P 500 index and stock options; \cite{chan2004net} for Hang Seng index options; \cite{shiu2010impact} for TAIEX index options and \cite{chuang2020impact} for VIX options. 
	\cite{kang2008information}  test the directional-learning hypothesis against limits-to-arbitrage and  conclude that  KOSPI 200 index options implied  volatilities  are largely driven by directional trading.  Further extending the net buying pressure metrics  to  distinguish between volatility-informed and directional-informed learning hypotheses, \cite{chen2017net} conclude that both directional and volatility motivated demand drive implied volatility changes in the TAIEX options market. This result is supported for the KOSPI 200 market by \cite{ryu2021impact}

	We contribute to this literature in several respects. First, previous studies  focus on established and regulated options markets, where sophisticated traders generally lead the market \citep{lakonishok2007option}. 
	We examine bitcoin options, which are traded in many different venues, most of them being self-regulated, and consequently there are concerns about market mature and price and volume manipulation  -- see \cite{makarov2020trading}, \cite{griffin2020bitcoin},  \cite{cong2020crypto} and others. 		{In 2019,  the US Securities and Exchange Commission (SEC) were persistently refusing applications to list bitcoin exchange-traded funds, their reason being that the market is not sufficiently mature.\footnote{See the Coindesk report: \href{https://www.coindesk.com/markets/2019/10/18/what-to-make-of-the-secs-latest-bitcoin-etf-rejection/}{What to Make of the SEC's Latest Bitcoin ETF Rejection.}} Our results could somehow alleviate regulator concerns about market immature by showing   Deribit bitcoin options have been aggregating volatility and directional information more efficiently since 2019.  However,  it does not help to alleviate concerns over
		manipulation. In fact, while spot bitcoin ETF applications are still not succeeding, the reason given nowadays is the presence of informed traders using private information to manipulate bitcoin prices.\footnote{See the Bloomberg report: \href{https://www.bloomberg.com/news/articles/2021-11-12/sec-rejects-vaneck-s-bitcoin-etf-application-to-trade-on-cboe}{SEC Rejects VanEck’s Bitcoin ETF in Latest Spot-Listing Snub}.}} 
	
	
	Secondly, the highly volatile nature of bitcoin  leads to compelling demand for options to manage both volatility and direction risk \citep{alexander2020bitmex, alexander2021optimal, deng2020minimum, scaillet2020high}. To meet these demands Deribit (and other bitcoin options platforms) offers extremely short-term 1-day and 2-day options, as well as one-, two- and three-weekly listings, plus the more standard quarterly expires. The very short-term maturities (1-day and 2-day) are unique to crypto options, and on the Deribit bitcoin platform they constitute almost 20\%  of the total volume. Therefore, we are able to disaggregate net buying pressure metrics and the subsequent regression results into different maturity buckets to investigate the type of informed trader operating at all points along the term structure of options maturities. 
	
	Thirdly,  whereas margin requirements on established stock and index options exchanges are directly linked to the underlying volatility,  Deribit and many other self-regulated bitcoin options platforms offer up 100X leverage. These markets also trade 24 hours a day, every day of the year.  This way they attract a broad participation from many uninformed speculators, not only from informed professional traders and arbitragers. 
	All these features are unique to bitcoin option markets, and they  could generate option trading motivations and consequent bitcoin price movements that are very different from those previously documented in established options markets.

	{Among other results, we find that both volatility and directional-motivated demands drive short and medium term OTM and DOTM options prices, whereas only directional trading drives prices of longer-term OTM calls and only volatility information drives prices of longer-term DOTM puts. How could bitcoin option traders benefit from understanding these driving force behind options of different moneyness?   Suppose a trader believes that the bitcoin market is about to enter a particularly volatile month ahead  -- perhaps as a result of heightened economic uncertainty, or because sentiment indicators based on twitter, reddit or other social media sites are sending mixed signals. Then the trader should enter a \href{https://www.investopedia.com/articles/active-trading/070914/risk-reversals-stocks-using-calls-and-puts.asp\#toc-what-is-risk-reversal}{bearish risk reversal strategy} by writing DOTM calls and using the premium to buy DOTM puts of the same maturity. The strategy would make a profit in the event that the trader's beliefs about volatility are borne out within the life of the options. }

	Our empirical results clearly confirm the limits-to-arbitrage hypothesis. On the two learning hypotheses, we find strong evidence of trades based on volatility information regardless of option moneyness categories,  but the directional-learning effect is less pronounced in ATM options. Directional-motivated demand is mainly significant in  out-of-the money (OTM) and deep-out-of-the money (DOTM) options. We propose the following economic reasons why there is more informed trading on privileged volatility information than on information about directional price changes:  (i) the  fundamental value of bitcoin and its return drivers (\cite{zhang2021downside}) are far from clear and   not widely acknowledged, so it is not easy to be informed by directional changes;\footnote{{As shown in the literature, bitcoin price could be driven by market sentiment and economic policy uncertainty (EPU),     see twitter of  \cite{shen2019does} and the economic policy uncertainty (EPU) of  \cite{demir2018does}.}} (ii) bitcoin spot and derivatives markets are highly segmented so price discovery can shift rapidly from one derivatives market to another; and (iii) bitcoin is very much more volatile than equities, so ATM options have unusually high volatility sensitivity, especially at longer maturities.

	{The unregulated less mature bitcoin options market differs from the well-developed S\&P 500 index options market in several aspects: (i) bitcoin options market trade much more short term options (less than 7-day)  that  have much greater market risk and are more sensitive to volatility than longer term options.  Additionally, bitcoin options call-put ratios is about 54:46   but about 4:6 for the S\&P 500 options. Those  implies bitcoin option traders are less risk averse than S\&P 500 option traders; (ii) bitcoin options have a much more symmetric smile. The slopes of S\&P 500 implied volatility curves become more negative with increasing fears of a stock market crash, while the negative left slope and positive right slope of bitcoin options smiles indicate that both positive and negative price jumps are expected; and (iii) bitcoin volatility risk premium   is consistently negative,     much larger in magnitude than that of S\&P 500 options.  This shows  bitcoin option market makers find it more difficult to hedge their inventory risk – due to the price jump risk and less liquidity – and hence charge a higher risk-premium.}

	To investigate  net buying pressure at different maturities,  we apply the methodology of \cite{chen2017net} to  three different subsets of   options according to their maturities. Given the volume of trading on extremely short-term options, we use  `short-term', `medium-term',  and `longer-term' to group options with maturities lying in the categories  $[1,7]$, $[8,21]$ and $\geq 22$ days. We select these buckets so that trading volume is approximately equal in each. Even though the volatility sensitivity increases with maturity, we find that volatility-informed traders prefer short-  and medium-terms  options. Nevertheless, volatility information does drive the prices of longer-term, DOTM puts. But it is directional trading that dominates prices of  longer-term OTM calls. Overall, longer-term options encompass less informed trades than short- and medium-term  options.

	Because we use hourly data we can also examine time-of-day patterns in the buying pressures on bitcoin options. A  recently emerging hour-of-day pattern shows that the strongest directional demand pressure comes during Asian trading times, whereas volatility-motivated demand is strongest while European markets are open.  {Quite interesting, although European trading time    has largest volatility-demand, it has less information.  ATM options are mainly driving by volatility-motivated demand during Asian   and US trading time.  This finding is in line with the volatility-transmission results of \cite{alexander2021volatility} who show that most volatility transmission occurs during US and Asian trading times. Also directional learning is only supported in OTM and DOTM option categories  during Asian   and US trading time slots.}

	The trading volumes on bitcoin options were very low during 2017 and 2018 so we do not report our results on these years  here but they are available on request.  By further disaggregating results year by year,  we find  that  volatility- and directional-informed traders only emerged after 2019. The scale of informed trading has continually increased since then, and we conclude that the Deribit market place is now rapidly evolving into a venue for the efficient aggregation and dissemination of both directional and volatility information.

	The remainder of this paper is organized as follows:  Section \ref{sec_deribit_market} describes the Deribit options market {and compares its implied volality curve to S\&P 500 index options market}; Section \ref{sec_nbp} uses hourly data aggregated from tick-level transactions to report different measures of net buying pressure;  Section \ref{sec_methods} summarises the regression models and relates parameter estimates to tests for the different hypotheses; Section \ref{sec_empirical} presents our  empirical results for options of different moneyness, maturities {and time-of-day}; and  Section \ref{sec_conclusion} concludes the paper. 

	\section{ {A Comparison of Bitcoin and S\&P 500 Options Markets} }\label{sec_deribit_market}
	%
	
	The online trading platform Deribit   was launched in the Netherlands in June 2016 but during 2021 it moved registration to the Republic of Panama, thus escaping the scrutiny of European derivatives regulators. The listed products  are perpetuals, fixed-expiration futures and options on only two USD rates,  versus bitcoin and ethereum  respectively.  At the time of writing,  Deribit currently accounts for about 80\% of bitcoin option trading volume across global exchanges, most of the other 20\% being traded on other self-regulated centralised exchanges such as OKEx, LedgerX and Huobi.\footnote{See reports: \href{https://analytics.skew.com/dashboard/bitcoin-options}{skew} and \href{https://www.newsbtc.com/news/deribit-is-dominating-bitcoin-options-with-70-market-share-heres-how-they-did-it/}{Deribit is Dominating Bitcoin Options With 78\% Market Share: Here’s How They Did It.}} 	{Due to the extreme volatility of bitcoin prices, both directional and volatility motivated demand could be present in these options.}
	
	
	%
	
	{S\&P 500 European index  options are margined and cash-settled in USD. The well-developed and regulated Chicago Board Options Exchange  opens at 8:30 a.m. central time and closes at 3:15 p.m. central time from Monday to Friday each week. Global trading after hours is also available from 7:15 p.m. to 8:15 a.m.  Monday to Friday.
		The bitcoin options  listed on Deribit are also European style, but they trade  continuously every day of the year -- the market never closes. They are margined and cash-settled  in bitcoin instead of a fiat currency such as USD.}  
	The  underlying  is the Deribit bitcoin index price (BTC) that is an equally weighted average of bitcoin USD spot prices on several leading spot exchanges such as Bitstamp, Gemini, Bitfinex, Bittrex,  Itbit, Coinbase, LMAX Digital and Kraken. Exchanges having invalid data or   delayed order book  are excluded in calculating the index price. The nominal value of each option is 1 BTC and settlement is at 8:00 UTC time (coordinated universal time).  The exchange lists a rich variety of  standard European calls and puts with  expiry dates ranging from 1-day to a maximum of 12 months. Starting in 2019, the demand from traders to hedge and speculate over a very short horizon led to the continual issuance of 1-day and 2-day options as well as 1-week, 2-week and (more recently) 3-week options.

	{ Table   \ref{tab_description_ttm_by_year} decomposes the total volume traded on all bitcoin options per year, using our own categorisation of these options into short-term (1 -- 7 days), medium-term (8 -- 21 days) and longer-term options ($\ge$ 22 days). As more new expires of short- and medium-term bitcoin options are introduced  the most liquid bitcoin options have less than seven days to maturity,   trading volume increasing from 19.17\%  in 2018 to 38.09\%   in 2021. As for S\&P 500 index  options, the share of weekly options (1 -- 7 days) increases from 21.86\% in 2017 to 28.01\%  in 2020 and medium-term options (8 -- 90 days) accounts for over 50\% of total volume. In most markets the prices of short-term option prices are more volatile than  prices of long term options, due to the gamma effect. Unlike S\&P 500 options traders,  bitcoin options traders focus mostly on short-term  options, which supports the view that bitcoin options traders are less risk averse than S\&P 500 options traders. Also, the net-buying-pressure literature on S\&P 500 usually focuses on medium term options,  because of this volume imbalance. By contrast, the much larger volume share of short-term options, which seems unique to  bitcoin options market, motivates us to include them in our empirical study. If the bitcoin market becomes less volatile over time, and attracts more institutional investors, we might expect a trading volume distribution closer to that of S\&P  500 index options.}
	
	\begin{table}[h!]
		\centering
		\caption{Deribit and S\&P 500 Option Trading Volume: Percentages by Maturities}
		\label{tab_description_ttm_by_year}
		\begin{tabular}{cccccc}
			\toprule
			Time to   Maturity & 2017 & 2018 & 2019 & 2020 & 2021 \\
			\toprule
			& \multicolumn{5}{c}{Deribit Bitcoin Option} \\
			1-7 days & 27.35\% & 19.17\% & 35.31\% & 42.19\% & 38.09\% \\
			8-21 days & 21.46\% & 14.41\% & 26.73\% & 27.18\% & 30.54\% \\
			$\geq $ 22 days & 51.18\% & 66.42\% & 37.96\% & 30.63\% & 31.38\% \\
			\toprule
			& \multicolumn{5}{c}{S\&P 500 Index Option} \\
			1-7 days & 21.86\% & 21.21\% & 23.43\% & 28.01\% &  \\
			8-90 days & 66.95\% & 67.43\% & 61.30\% & 54.33\% &  \\
			$\geq $ 91 days & 11.19\% & 11.37\% & 15.27\% & 17.66\% & \\
			\toprule
		\end{tabular}
		\floatfoot{{Note. The figures above are from millions of tick-by-tick Deribit bitcoin options  trades from 1 January 2017 to 28 July 2021 accessed via the Deribit application program interface, which no longer provides historical data but our sample  can be accessed from \href{https://www.deribit.com/pages/information/tardis}{Tardis} or  \href{https://www.coinapi.io/}{CoinAPI}. The sample stops after July 2021 because there are data only for very short maturity trades. 	Daily frequency S\&P 500 index options data is from Wharton Research Data Services. It has 8,688,408 entries, containing 4,362,860 (50.21\%) calls  and 4,325,548 (49.79\%) puts, but the data is only available  until 2020.}}
	\end{table}

	\newpage
	In addition to option expiration date, strike (in USD) and type (European call or put),  our bitcoin option tick-by-tick data includes details about the trading direction (i.e. whether buyer or seller initiated), the implied volatility, the option price (in BTC),  the traded amount (in BTC), the underlying bitcoin index price (in USD)  and the timestamp in milliseconds. The full data-set contains 3,431,071 entries consisting of 1,834,499 (53.47\%) call options and 1,594,420 (46.47\%) put options. For 2,152 (0.062\%) of trades the option type was missing. The Deribit exchange imposes  an upper bound of 500\% and a lower bound of 0\% for their implied volatility quotes due to concerns of data outliers and illiquidity. This way another  2,447 (0.71\%) of the full data-set were excluded. The cleaned final data-set  contains 3,427,160 entries consisting of 1,833,838 (53.51\%) calls, and 1,593,322 (46.49\%) puts.

	Options of different  strikes $K$ and maturities $T$ are classified by  their Black-Scholes deltas\footnote{{Other authors use $K/S,  \ \log(K/S)$ and $K/S e^{-rT}$ to measure moneyness. We also tried these alternatives, but found virtually no difference in our qualitative conclusions. The S\&P 500 options data already contains Black-Scholes  $\Delta$ and since this metric is also standard in the net buying pressure literature, we also use $\Delta$.}}, viz.
	\begin{align}
		\Delta^C _t(F_t,\sigma|K,T)&=\Phi\left[ \frac{\ln \left( F_t/K \right) +0.5\sigma ^2\tau}{\sigma \sqrt{\tau}} \right] \quad \mbox{and} \quad  \Delta^P _t(F_t,\sigma|K,T) =\Delta^C _t(F_t,\sigma|K,T) -1,
	\end{align}
	where $ \Phi[\cdot] $ is the standard cumulative normal distribution function, $S_t$ is the underlying price (USD), $F_t=S_te^{r\tau}$ is the forward price (USD),  $ K$ is the strike price (USD), $\sigma$ is the annualised volatility, $\tau = T - t$ is the residual time to maturity in years and we use $C$ for a call and $P$ for a put.\footnote{{We only use the Black-Scholes model to recover implied volatilities and deltas in order to group options into different moneyness categories. Of course, a geometric Brownian motion is not suitable to model the bitcoin underlying price --  stochastic volatility models such as those in \cite{tiwari2019modelling} or \cite{siu2021bitcoin} would be preferable.}} Following  \cite{bollen2004does}, most papers in the literature set $\sigma$ equal to the \textit{realised} volatility, calculated as the annualised  square root of the average sum of squared log returns over the most recent $n$-days.\footnote{The choice for $n$ differs according to the data frequency. For instance, \cite{bollen2004does} use 60 days for their daily data, and \cite{kang2008information} and \cite{chen2017net} both use 60 days for their 5-minute data.} We use a 15-day realised volatility based on bitcoin daily log returns, the same $\sigma$ for all options. {Of course, using the option's own implied volatility for dynamic delta hedging but using a physical measure of volatility for calculating the option's moneyness is a standard practice in this literature -- it is attractive because it avoids any endogeneity concerns in our regressions where the dependent variable  is implied volatility. Besides, it is important to note that (i) we only use this to classify options into five moneyness categories, and (ii) our results are robust to alternative choices for $\sigma$}.\footnote{Using a 15-day realised volatility is, of course, ad hoc. So we explored some alternatives for $\sigma$ which slightly change the way that options are assigned to  moneyness categories. We analysed results for a 30-day realised volatility, using the option's own implied volatility, and using a simpler definition of moneyness as the ratio of the underlying price to the strike of the option, as in \cite{ryu2021impact}. We found virtually no difference in the qualitative conclusions to be drawn from our analysis. Detailed results are available from the authors on request. } Now we use the option deltas to allocate them into one of five moneyness categories, as follows:\footnote{Options with absolute delta below 0.02 or above 0.98 are excluded because of the distortions due to stale prices and illiquidity. }   
	\begin{itemize}
		\item Deep-out-of-the-money (DOTM): $ 0.02 <|\Delta |\le 0.125$,
		\item Out-of-the-money (OTM): $ 0.125<|\Delta |\le 0.375$,
		\item At-the-money (ATM): $ 0.375<|\Delta|\le 0.625$,
		\item In-the-money (ITM): $ 0.625<|\Delta|\le 0.875$,
		\item Deep-in-the-money (DITM): $0.875< |\Delta |\le 0.98$. 
	\end{itemize} 
	
	\begin{table}[h!]
		\centering \small
		\caption{Trading Volume Across Five Moneyness Categories}
		\label{tab_moneyness_volume_byyear}
		\begin{threeparttable}
			\begin{tabular}{ccccccccccc}
				\toprule
				& \multicolumn{2}{c}{2017} & \multicolumn{2}{c}{2018} & \multicolumn{2}{c}{2019} & \multicolumn{2}{c}{2020} & \multicolumn{2}{c}{2021} \\
				& Call & Put & Call & Put & Call & Put & Call & Put & Call & Put \\
				\toprule
				& \multicolumn{10}{c}{Deribit Bitcoin Option} \\
				DOTM $(\%)$ & 2.86 & 10.34 & 18.44 & 7.75 & 14.62 & 13.83 & 12.74 & 12.78 & 13.71 & 13.43 \\
				OTM $(\%)$ & 13.93 & 16.85 & 21.66 & 13.24 & 21.12 & 20.15 & 21.74 & 21.41 & 22.61 & 23.01 \\
				ATM $(\%)$ & 26.27 & 9.10 & 16.24 & 7.73 & 12.86 & 9.11 & 14.39 & 11.04 & 12.87 & 9.74 \\
				ITM $(\%)$ & 14.47 & 2.13 & 7.20 & 4.05 & 4.46 & 2.17 & 3.38 & 1.28 & 2.54 & 1.12 \\
				DITM $(\%)$ & 3.53 & 0.53 & 1.61 & 2.08 & 1.16 & 0.52 & 1.04 & 0.20 & 0.71 & 0.25 \\
				Total Prop. $(\%)$ & 61.06 & 38.94 & 65.15 & 34.85 & 54.22 & 45.78 & 53.29 & 46.71 & 52.44 & 47.56 \\
				\# T.C. (Million) & 0.02   &  0.02 &    0.21&    0.09  &    0.70&    0.60 &    2.36&    2.08&    1.38&    1.32\\
				T.V. ($\$$Billion) & 0.16 & 0.10 & 0.93 & 0.50 & 4.27 & 3.60 & 23.59 & 20.68 & 53.99 & 48.97 \\
				\toprule
				& \multicolumn{10}{c}{S\&P 500   Index Option} \\
				DOTM $(\%)$ & 9.28 & 25.65 & 10.91 & 21.57 & 9.04 & 23.89 & 9.04 & 21.58 &  &  \\
				OTM $(\%)$ & 13.99 & 22.20 & 15.38 & 22.21 & 14.18 & 21.27 & 13.45 & 22.97 &  &  \\
				ATM $(\%)$ & 12.93 & 11.00 & 11.70 & 12.62 & 14.27 & 12.76 & 13.51 & 13.44 &  &  \\
				ITM $(\%)$ & 2.80 & 1.21 & 2.25 & 2.35 & 2.55 & 1.29 & 2.86 & 2.00 &  &  \\
				DITM $(\%)$ & 0.79 & 0.15 & 0.48 & 0.53 & 0.48 & 0.27 & 0.67 & 0.47 &  &  \\
				Total Prop. $(\%)$ & 39.79 & 60.21 & 40.72 & 59.28 & 40.52 & 59.48 & 39.54 & 60.46 &  &  \\
				\# T.C. (Million) & 88.37 & 133.70&  113.60 &  165.40 &  88.45 &  129.81 &  82.30 &  125.86   &  &  \\
				T.V. ($\$$Billion) & 168.95 & 179.76 & 311.98 & 439.71 & 289.06 & 331.36 & 537.93 & 751.90 &  &    \\
				\toprule 
			\end{tabular}
			\begin{tablenotes}
				\item \small
				Note. {The total number of traded   contracts (T.C.) and}  USD trading volume (T.V.) by year and the proportion traded in each moneyness category, are decomposed into calls and puts. {The notional value per bitcoin option contract is   1 bitcoin and one point of S\&P 500 index option is 100 USD. }
			\end{tablenotes}
		\end{threeparttable}
	\end{table}
	
	\normalsize
	Table \ref{tab_moneyness_volume_byyear} reports {the total number of traded   contracts and} USD trading volumes     on all options by year (in USD billion, last line) and decomposes this into  proportions traded in each moneyness category above, and into calls and puts within each category.  {The  annual  number of traded contracts of S\&P 500   options is quite stable, siting at roughly 200 million per year, except 279 million in 2018. The bitcoin options total USD trading volume increases 400-fold, from only 0.26 USD billion  in 2017 to 102.96 USD billion  in 2021 and  the total number of traded contracts increases 111-fold from 0.04 million in 2017 to 4.4 million in 2020.  Another difference is that the S\&P 500  call-put ratio is almost identical, around 4:6 from 2017 to 2020, while traders prefer bitcoin calls instead of puts and  the  bitcoin call-put ratio changed significantly, from about 65:35 in 2018 to about 54:46 after 2019. This shows that the bitcoin options market   has less risk-averse investors than the S\&P 500 options market, which is led by sophisticated institutional traders who demand out-of-the-money puts to hedge downside and crash risk, see \cite{kelly2016too}.}

\begin{figure}[h!]
		\includegraphics[trim=3.5cm 0cm 4cm 1.3cm, clip=true, width=0.9\linewidth]{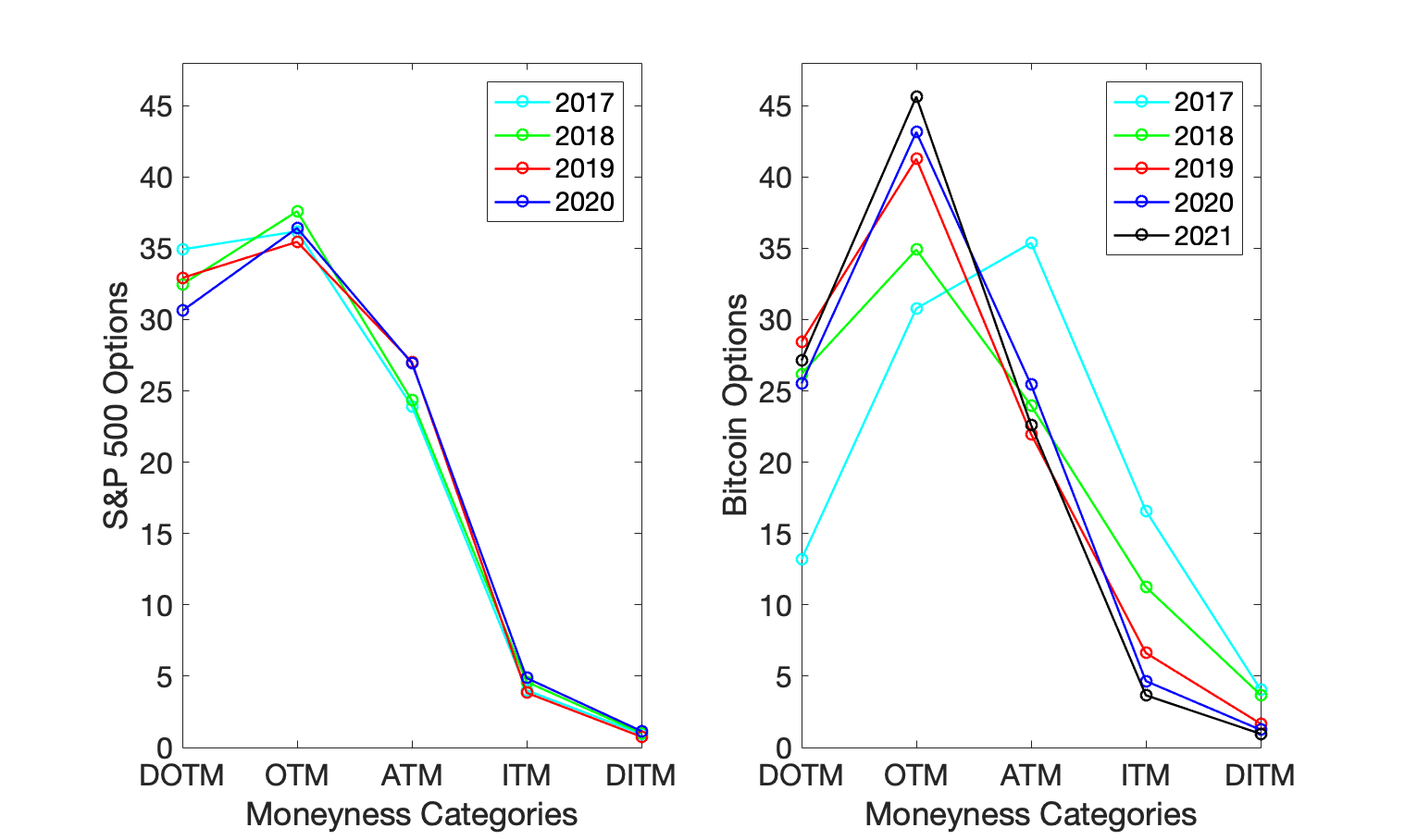}
		\caption{S\&P 500 and Bitcoin Options Volume Percentage by Moneyness Categories}
		\label{moneyness_compare}
		\floatfoot{Note. The left and right panels plot the trading volume percentages by moneyness categories of  S\&P 500     and bitcoin options accordingly. The pattern  of S\&P 500   options   is quite stable    where OTM, DOTM and ATM are the most traded contracts with shares about 32\%, 35\% and 25\% respectively.  After 2019, the shares of bitcoin options are converging to a pattern (even in magnitude) very similar to that of the S\&P 500 options.}
	\end{figure}

	{The two options markets demonstrate a similar   volume distribution for different moneyness categories, where   OTM, DOTM and ATM options are the most traded contracts and  trades on ITM and DITM options are seldom -- less than 4\% for both bitcoin and   S\&P 500 options.   This is to be expected, because the call (put) option price increases rapidly as the strike decreases (increases). For instance, in the bitcoin options market  during 2021 (January to July)  total volume reached 52.44 USD billion for calls and 47.56 USD billion for puts, and 3.25\% of the total volume was traded on ITM and DITM calls and even less, 1.37\%, on puts.  
		The left and right panels of  Figure \ref{moneyness_compare} depict  how the percentages of total trading volume in the five moneyness categories have changed over time for S\&P 500     and bitcoin options respectively. The pattern  of S\&P 500   options   is quite stable    where OTM, DOTM and ATM are the most traded contracts with shares about 32\%, 35\% and 25\% respectively.  Although much variability shown    in 2017 and 2018, where OTM bitcoin options increase from   30\% in 2017 to over 45\%  in 2021,  the shares of bitcoin options are converging together during 2019-2021 to similar pattern (even in magnitude) as for S\&P 500 options.  This indicates the bitcoin options market only really started maturing in 2019, so more recent data are likely to be more reliable and not distorted by stale prices or low trading  volumes.
	}

	{Now we compare and contrast  the implied volatility curves of  bitcoin  and S\&P 500 options,  interpreting the implications of their similarities and differences for bitcoin option traders. As shown in Table \ref{tab_btc_sp_iv}, bitcoin realised volatility (RV) and implied volatility (IV) are both much larger (around 4-times in 2018 and 2019) than the corresponding metrics for S\&P 500 options -- although, in line with our observation from  Tables \ref{tab_btc_sp_iv} and \ref{table_iv_slope}, a  trend of decreasing volatility in bitcoin is perceived as the options market matures. The volatility risk premium, as measured by the realised-implied volatility spread (VS = RV-IV) is   consistently negative and much larger in magnitude for bitcoin options. This is not surprising because bitcoin has a great risk of price jumps and some options have low liquidity, making it more difficult for bitcoin options market makers to hedge their inventory risk. In compensation, for holding this risk they   can charge a higher volatility risk premium than S\&P 500 options market makers.}

	\begin{table}[h!]
		\caption{{Realised Volatility, Implied Volatility and Volatility Spread}}
		\label{tab_btc_sp_iv}
		\begin{tabular}{cccc|ccc|c}
			\toprule
			& \multicolumn{3}{c|}{Bitcoin Option} & \multicolumn{3}{c|}{S\&P 500 Index Option} & \\
			\toprule
			Year & RV         & IV        & VS        & RV           & IV           & VS          &        IV Correlation         \\
			\toprule
			2017&	0.93&1.05	&-0.12	&0.08&	0.17&	-0.09&	0.02 \\
			2018 & 0.85       & 0.97      & -0.12      & 0.17         & 0.20         & -0.03        & -0.02          \\
			2019 & 0.65       & 0.80      & -0.15      & 0.15         & 0.19         & -0.04        & 0.07           \\
			2020 & 0.64       & 0.76      & -0.12      & 0.32         & 0.31         & 0.002        & 0.17      \\
			\toprule    
		\end{tabular}
		\floatfoot{{Note. The implied volatility (IV)     is  calculated as the average of all trades in each year and the  {realised} volatility is calculated as the annualised  square root of the average sum of squared log returns over the most recent 15-days. The realised-implied volatility spread (VS=RV-IV) measures the risk-premium charged by market markers for providing liquidity and  compensating  inventory risk. }}
	\end{table}
	

	\begin{table}[h!]
		\centering
		\caption{{Implied Volatility Curve}}
		\label{table_iv_slope}
		\begin{tabular}{cccc|ccc}
			\toprule
			& \multicolumn{3}{c|}{Deribit Bitcoin Option} & \multicolumn{3}{c}{S\&P 500 Index   Option} \\
			Year & Level & Left Slope & Right Slope & Level & Left Slope & Right Slope \\
			\toprule
			2017 & 0.996 & -0.035 & 0.045 & 0.108 & -0.312 & -0.200 \\
			2018 & 0.868 & -0.026 & 0.050 & 0.142 & -0.296 & -0.166 \\
			2019 & 0.716 & -0.032 & 0.023 & 0.142 & -0.285 & -0.180 \\
			2020 & 0.618 & -0.081 & 0.039 & 0.247 & -0.284 & -0.176 \\
			mean & 0.799  &	-0.044  &	0.039  &	0.160  &	-0.294 & 	-0.180 \\		
			\toprule
		\end{tabular}
		\floatfoot{Note.  The {level}    is  calculated as the average of all ATM calls and puts in each year, regardless of expiration. The  {left} slope   is defined as the ratio of   difference between IVs of category 2   and  category 3 options to the IV level, while the  {right} slope is    the ratio of difference between IVs of category 3 and category 4 options to the   IV level. Category 2: ITM calls and OTM puts. Category 3: ATM calls
			and ATM puts. Category 4: OTM calls and ITM puts.}
	\end{table}

	\begin{figure}[h!]
		\includegraphics[trim=2.7cm 4.2cm 2.7cm 3.9cm, clip=true, width=1\linewidth]{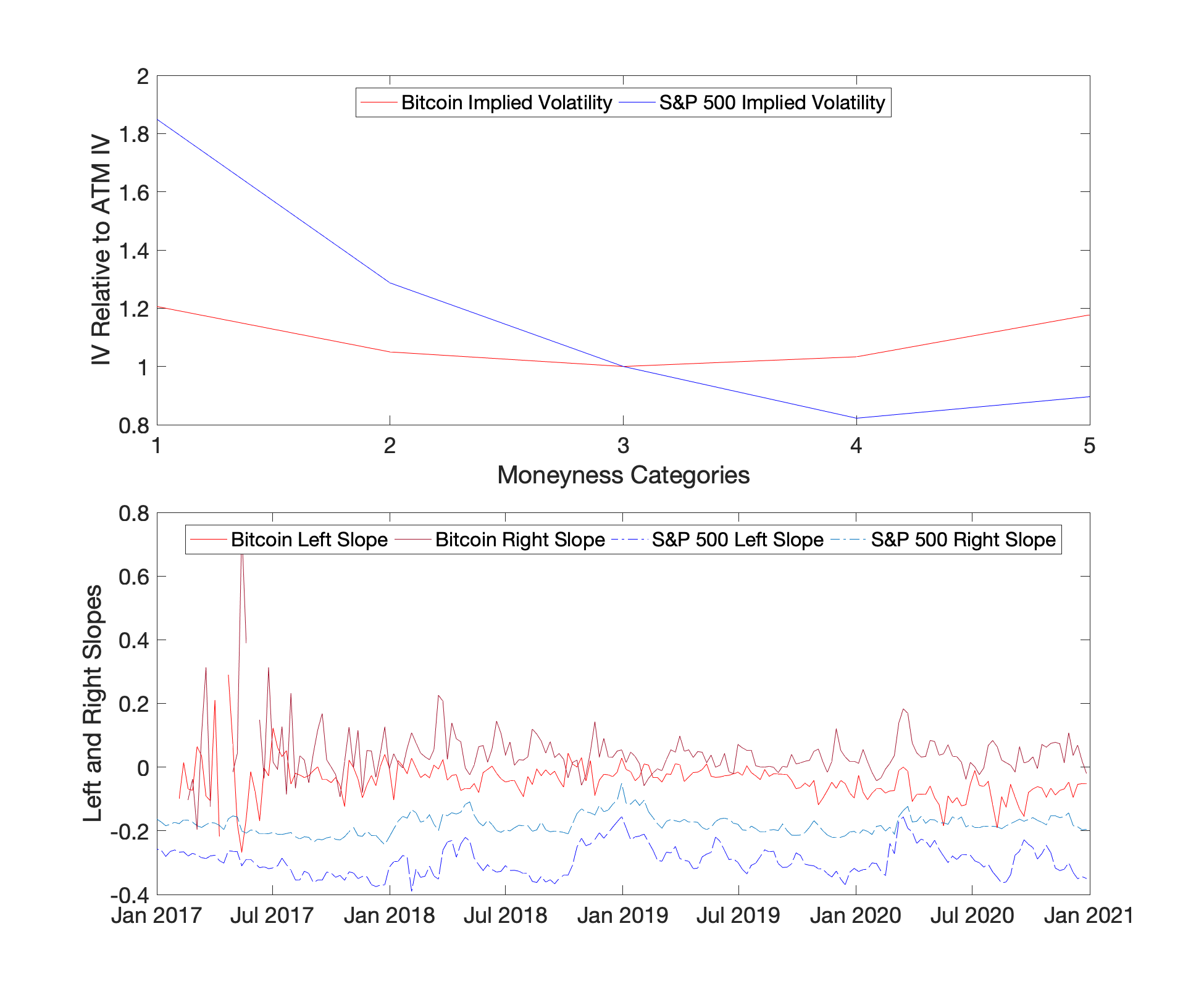}
		\caption{{Relative-to-ATM Implied Volatility Smile and Slopes}}
		\label{btc_sp500_slope}
		\floatfoot{Note. To make the much higher bitcoin IVs comparable to  S\&P 500 options, we scale   implied volatilities of all categories by its own  ATM IV. Top panel:   it  plots    implied volatilities relative to  ATM options' IV  for the five categories. 
			Bottom panel: it shows the weekly average of left and right slopes  from 2017 to 2020.    The  {left} slope   is defined as the ratio of   difference between IVs of categories  2   and    3 options to ATM IV, while the  {right} slope is    the ratio of difference between IVs of categories  3 and   4 options to ATM   IV.  }
	\end{figure}
	{To characterise the implied volatility curve (IVC), we group implied volatilities (IV) as follows: Category 1: DITM calls and DOTM puts. Category 2: ITM calls and OTM puts. Category 3: ATM calls
		and ATM puts. Category 4: OTM calls and ITM puts. Category 5: DOTM calls and DITM puts. Then, to make the much larger     bitcoin  and  S\&P 500 options IVs on a comparable scale, we divide   implied volatility for all categories by   the implied volatility of   ATM options in Category 3. Additionally,  following the literature, we use the terminology \textit{IV level}       to denote the average IVs of ATM calls and puts,  \textit{{left slope}} is the difference between  average IVs of Category 2   and  Category 3 options, divided by the IV level, and  \textit{right slope}  is    the difference between average IVs of Category 3 and Category 4 options, divided by the   IV level. Table \ref{table_iv_slope} reports the results and Figure \ref{btc_sp500_slope} depicts them. They  demonstrate several importance differences between bitcoin and S\&P 500 options IVs. }	

	{First, from Table \ref{tab_btc_sp_iv}, the bitcoin IV level is roughly five times  larger   than that of S\&P 500 options, and only in bitcoin  there is a downward trend in average IV level -- it decreases year by year as   the bitcoin spot and options markets mature over time. 	Second, as seen in both the Table \ref{table_iv_slope} and Figure \ref{btc_sp500_slope} the S\&P 500 options IVC is much steeper than that of   bitcoin  options and its typical  `skew' shape  is not mirrored by bitcoin options, which have a much more symmetric smile. This is not surprising given their statistical properties: the S\&P 500 realised skewness (-0.732) is negative and about ten times the size  of  the bitcoin realised skewness (-0.0715), and the realised kurtosis is greater in bitcoin.  \cite{bakshi2003stock} show that the IVC  is shaped by the  skewness and kurtosis in the physical measure:    
		the more negative the skewness,  the steeper the implied volatility curve.  
		The left and right slopes of S\&P 500 options are both negative, downward-sloping and rather stable across time, around -0.3 and  -0.2 accordingly. In contrast,  the left slope of bitcoin implied volatility curve is negative with four years average -0.044 and the right slope is positive  with four years average 0.039.  The bottom panel of  Figure \ref{btc_sp500_slope} plots the evolution of the left and right slope measures over time.   The slopes of S\&P 500 implied volatility curves become more negative with increasing fears of a stock market crash, while the negative left slope and positive right slope of bitcoin smiles indicate that both positive and negative price jumps are expected.} 
	
	{	Clearly, bitcoin and the S\&P 500 index are driven by different  underlying jump-diffusion stochastic   processes. For instance,  \cite{yan2011jump} proves that the implied volatility slope is approximately proportional to the average jump size and jump intensity of the underlying   price process, and \cite{scaillet2020high} found  a total 124 large jumps in the bitcoin price, containing 70 positive  and 54 negatives jumps, during the period 26 June 2011 to 29 November  2013 alone. }

	\section{{The Information Content of Bitcoin Options}}\label{sec_nbp}	
	
	{We begin this section by examining the relationship between the change in bitcoin option trading volume $\Delta v$ or the aggregated net-buying-pressure (i.e. the order imbalance)  and one-period-ahead bitcoin spot return or volatility, either realised and implied. We perform three separate sets of regressions that test the predictive power of volume and order-imbalance information metrics for bitcoin spot and/or options prices. First, to measure order imbalance over a time  period between time $t-1$ and  time $t$, we define the order imbalance  $N_t$ as the difference between the aggregated volumes on all buyer-initiated trades and seller-initiated   trades, each multiplied by the absolute value of the options  delta, i.e.:}
	
	{	\begin{align}\label{eq_all_nbp_calculation}
			N_{t}=\sum_{i=1}^{m_{t}} {B_{i,t} \times \left| \Delta _{i,t} \right|}-\sum_{i=1}^{n_{t}} {S_{i,t}\times}\left| \Delta _{i,t} \right|,
		\end{align}
		where  $ B_{i,t}$ is the volume on the $i^{th}$ buyer-initiated trade during the interval between time $t-1$ and time $t$, and  $ S_{i,t}$ is the equivalent quantity for  seller-initiated trading volume, and $|\Delta _{i,t}|$ is the absolute value of the delta of   the option that is traded. This way, the  buying side is aggregated over $m_{t}$, the number of of buyer-initiated trades  between times $t-1$ and  $t$, and the selling side is summed over $n_{t}$, the corresponding number for seller-initiated trades. Using these data we run the following regressions:}
	\begin{align}\label{eq_return_regre}
		{x_t = \gamma_0 + \gamma_1\cdot  x_{t-1} + \gamma_2 \cdot y_{t-1}, \qquad x\in\{r, \Delta IV,  \Delta RV\}, \quad 
			y \in \{ \Delta v, N\}}.
	\end{align}

	\begin{table}[h!]\centering
		\caption{{Information Content of Option Trading}}
		\label{tabeq_return_regre}
		\begin{center}
			\begin{tabular}{ccccc|ccc}
				\toprule
				&  & \multicolumn{3}{c|}{Volume, $\Delta v$ } & \multicolumn{3}{c}{Aggregate Order Imbalance, $N$} \\
				&  & $\gamma_0$ & $\gamma_1$ & $\gamma_2$ & $\gamma_0$ & $\gamma_1$ & $\gamma_2$ \\	
				\toprule
				& 1-Hour & $0.0001^{***}$ & $-0.053^{***}$ & $0.27^{***}$ & $0.0001^{***}$ & $-0.053^{***}$ & $8.43$e-7 \\
				Return & 1-Day & $0.002^{***}$ & $-0.084^{***}$ & $0.51^{***}$ & $0.0016^{***}$ & $-0.086^{***}$ & $-1.38$e-6 \\
				& 5-Days & $0.016^{***}$ & $-0.025$ & $0.068$ & $0.018^{***}$ & $-0.0027$ & $-6.08$e-6 \\
				\toprule
				& 1-Hour & $-0.0019^{***}$ & $-0.42^{***}$ & $4.27^{***}$ & $-0.0021^{***}$ & $-0.42^{***}$ & $6.36$e-5$^{***}$ \\
				IV & 1-Day & $-0.0036^{***}$ & $-0.067^{***}$ & $3.32^{***}$ & $-0.0055^{***}$ & $-0.11^{***}$ & $3.96$e-5$^{***}$ \\
				& 5-Days & $-0.0077$ & $0.14^{***}$ & $1.73^{***}$ & $-0.010^{***}$ & $0.057$ & $3.02$e-6 \\
				\toprule
				& 1-Hour & $-1.02$e-5$^{***}$ & $0.034^{***}$ & $0.18^{***}$ & $-1.49$e-5$^{***}$ & $0.031^{***}$ & $4.35$e-7$^{***}$ \\
				RV & 1-Day & $-0.0007^{***}$ & $0.41^{***}$ & $0.30^{***}$ & $-0.0012^{***}$ & $0.39^{***}$ & $9.94$e-6$^{***}$ \\
				& 5-Days & $-0.0078$ & $0.35^{***}$ & $-0.48$ & $-0.014^{***}$ & $0.35^{***}$ & $2.89$e-5$^{***}$ \\
				\toprule
			\end{tabular}
		\end{center}
		\begin{tablenotes}
			\item \small 
			Note. This table reports the regression  result $x_t = \gamma_0 + \gamma_1\cdot  x_{t-1} + \gamma_2 \cdot y_{t-1}$. Here $x_t$ is the bitcoin price return or $\Delta IV$ or $\Delta RV$; and $y_t$ is   the change of    bitcoin options trading volume $\Delta v_t$ or the order imbalance measured by $N_t$. The   $^{***}$ indicates   1\% significance level.  
		\end{tablenotes}
	\end{table}
	{The results are displayed in Table \ref{tabeq_return_regre}. Several highly significant  estimates for $\gamma_2$  confirm that hourly trading volume has a significant positive effect on returns and realised volatility over the next hour, and a similar positive causation is clear from our regressions of daily returns or realised volatility on the previous day’s trading volume. Options trading volume has an even greater effect on future implied volatility, where the significance in implied volatility (and therefore also option prices) persists even for the next five days. 
		Hourly and daily net buying pressure also have a highly significant effect on both realised and implied volatilities.} 
	
	{In contrast to trading volume, order imbalance  during the hourly, daily or 5-day interval has no predictive power for returns over the next interval.  This is understandable, and supported by our empirical analysis later in this paper, where we show that order imbalance  is  motivated more by volatility information than information about the direction of the underlying price, i.e about the sign and size of the return. And in supporting of later results, order imbalance does have predictive power for implied and realised volatilities. So overall, the results in Table \ref{tabeq_return_regre} suggest that bitcoin options traders may indeed possess superior, if short-lived, information
		about the future direction of bitcoin spot and option prices that would allow them to generate
		abnormal returns by developing a suitable trading strategy for bitcoin options.}

	{Section \ref{sec_deribit_market} has demonstrated that the bitcoin options market is less mature than the S\&P 500 options market. Spot prices have more frequent positive as well as negative jumps, and bitcoin option traders appear less risk averse than S\&P 500 options traders. Hence, we might expect different trading motives to drive the bitcoin options implied volatility curve compared with other  well-regulated, more mature markets such as the S\&P 500 options market. 
		To investigate these motives we start by following} \cite{bollen2004does} to define the  net buying pressure at time $t$ in moneyness   $k$, denoted $A^k_t$ as the difference between   buyer-initiated trades and seller-initiated   trades multiplied by the absolute value of the options  Black-Scholes delta.  More precisely, we also divide net buying pressure to separate that on calls from that on puts, therefore defining it as:
	\begin{align}\label{eq_nbp_calculation}
		A_{j,t}^{k}=\sum_{i=1}^{m_{jk,t}} {B_{ij,t}^{k}\times \left| \Delta _{ij,t}^{k} \right|}-\sum_{i=1}^{n_{jk,t}} {S_{ij,t}^{k}\times}\left| \Delta _{ij,t}^{k} \right|,
	\end{align}
	where   $ j\in \left\{ \mbox{Call, Put}  \right\} =\left\{C, P\right\}$, $ k\in \left\{ \mbox{DOTM, OTM, ATM, ITM, DITM} \right\} $ and $ B_{ij,t}^{k} $ (resp. $ S_{ij,t}^{k} $) is the  buyer-initiated (seller-initiated) trading volume  on the $i^{th}$ option of type $j$ in moneyness category $k$. This is aggregated over all buyer- or seller-initiated trades  between time $t-1$ and  time $t$, weighted by the absolute value of $ \Delta _{ij,t}^{k} $, the delta of the option traded. This way, the  buying pressure is aggregated over $m_{jk,t}$, the number of of buyer-initiated trades on an option for type $j$ in moneyness category $k$, between times $t-1$ and  $t$, and the selling pressure is summed over $n_{jk,t}$, the corresponding number for seller-initiated trades.

	The original metric for net buying pressure defined above mixes directional-informed and volatility-informed trading. Therefore, following \cite{chen2017net} (also recently used in \cite{ryu2021impact}), at each moneyness level $k$ we separately measure the directional-motivated demand for calls and puts, calculated as:
	\begin{equation}\label{eq_direction_NBP}
		D_{C,t}^{k}=\frac{A_{C,t}^{k}-A_{P,t}^{k}}{2}, \quad   \mbox{and} \quad D_{P,t}^{k}=\frac{A_{P,t}^{k}-A_{C,t}^{k}}{2}. 
	\end{equation}
	Note that $D_{P,t}^{k} = - D_{C,t}^{k} $ by definition. Equation   \eqref{eq_direction_NBP} captures directional trading volume in that an anticipated rise (fall) in the underlying price would induce a positive (negative) net demand for calls and negative (positive) net demand for puts.  That is, when directional traders dominate the volume in moneyness category $k$, then  $D_{C,t}^{k} > 0$ indicates an anticipation of a rise in the underlying price, whereas $D_{C,t}^{k} < 0$ means the price is anticipated to fall. Also following \cite{chen2017net} the volatility-motivated demand (which is the same for  calls and puts) is defined as:
	\begin{equation}\label{eq_vol_NBP}
		V_{t}^{k}=\frac{A_{C,t}^{k}+A_{P,t}^{k}}{2}.
	\end{equation}
	Equation \eqref{eq_vol_NBP} captures anticipated increasing (decreasing) in implied volatility by volatility-informed traders, which would induce positive (negative) net buying pressure on both calls and puts. That is, when volatility traders dominate the volume in moneyness category $k$, $V_{t}^{k}>0$ when these traders expect implied volatility  to increase and  $V_{t}^{k}<0$ if they expect implied volatility to decrease.\footnote{{We could also scale the directional and volatility-motivated demands by the total delta-weighted volume $${TV}^k_t =  {\sum_{j\in \{call, put\}} \sum_{i=1}^{m_{jk,t}} {B_{ij,t}^{k}\times \left| \Delta _{ij,t}^{k} \right|} + \sum_{i=1}^{n_{jk,t}} {S_{ij,t}^{k}\times}\left| \Delta _{ij,t}^{k} \right|}$$ and define the relative net-buying-pressures as
			$$
			\overline{D}_{C,t}^{k} =\frac{D_{C,t}^{k}}{{TV}^k_t}, \qquad 
			\overline{D}_{P,t}^{k} =\frac{D_{P,t}^{k}}{{TV}^k_t}, \qquad
			\overline{V}^k_t = \frac{V_{t}^{k}}{{TV}^k_t}.	 
			$$
			The empirical results are available on request, but not reported here because they are qualitatively  unchanged. We thank an anonymous referee for bring this robustness check to our attention. }}

	We obtain an hourly time series for net buying volume, net buying pressure, and directional- and volatility-motivated demand. For net buying volume we aggregate volumes on all buyer-initiated trades during the 1-hour interval, and similarly for the seller-initiated trades, and then take the difference.  For the other measures, we aggregate them  using  formulae \eqref{eq_nbp_calculation}-\eqref{eq_vol_NBP}. The previous literature uses  different time intervals  for aggregation:  daily frequency  in \cite{bollen2004does}, five-minutes in \cite{kang2008information} and \cite{chen2017net}     and   1-minute  in \cite{ryu2021impact}.

	For illustration, Figure \ref{fig_time_series_nbp} contains six time series plots of directional- and volatility-motivated net buying pressure during the last three years of the sample,  i.e. 1 January 2019 to 28 July 2021.   The left panel exhibits the first part from 1 January 2019 to 30 April 2020 and -- on a different scale --  the right panel exhibits the second part, from 1 May 2020 to 28 July 2021.   The upper plots depict the net buying and  selling volumes during the day, and the remaining plots depict the dynamics of $V_t^k$ and $D_t^k$ for $k =$ DOTM, OTM, ATM, ITM and DITM. We observe that the time series increase dramatically and are much more volatile after 1 November 2020 when the bitcoin price rise from 10,000 USD to more than 60,000 USD in the space of a few months. 

	\begin{figure}[htp]
		\centering
		\includegraphics[trim=0.2cm 1cm 0.2cm 0cm, clip=true, width=1\linewidth]{fig/ts_nbp.png}
		\caption{The Dynamics of Directional- and Volatility-Motivated Net Buying Pressure}
		\label{fig_time_series_nbp}
		\floatfoot{Note. Net buying/selling volume on Deribit bitcoin options (top) and  volatility-motivated (red) and directional-motivated (blue) net buying pressure for options in different moneyness categories, measured  in USD million. We only display the series for  calls, because the respective measure for puts is either the same (volatility-motivated) or the opposite sign  (directional-motivated).  The left panels are from May 2019 to April 2020;  the right panels are from May 2020 to July 2021. The hourly measures are averaged over each day and the plots exhibit these daily averages.}
	\end{figure}

	Table \ref{tab_moneyness_volume_NBP} reports the average net buying volume, and different measures of net buying pressure on calls and puts, where the hourly data are now averaged over each year in the sample. All metrics increase dramatically over time. For instance, the annual average  ATM net buying pressures shot up from 455 USD in 2017 to  94,818 USD  in 2020. Traders consistently sell  DOTM and OTM options and  buy ATM, DITM and ITM options over the entire period, except for the small net selling of ITM and DITM put options in 2021. As a result, the net buying pressure on ITM and DITM options is exceptionally large. Even though fewer such options are traded, as noted from Table \ref{tab_moneyness_volume_byyear} above, the buyer-initiated trades far outweigh the seller-initiated trades, particularly on calls. On average during 2020, the net buying pressure on ITM calls is 109.48 USD million, which is about 2/3rds of the buying pressure on ATM calls, and the average net buying pressure on DITM calls is almost the same as for ITM  calls. This is much larger than is seen in standard markets. For instance, traders on  S\&P 500 index options mostly buy DOTM and OTM puts for protection and sell DITM and ITM calls for premiums in \cite{kang2008information}.  The unusual demand for ITM and DITM calls might be induced by   unusually large directional-motivated demand, probably based on the prediction of future price bubbles. Such bubbles have been  frequent, even  during bitcoin's short history  \citep{cheah2015speculative} and predictions of further bubbles are commonly and consistently reported in the media.

	\begin{table}[h!]
		\centering \small
		\caption{Average Hourly Net Buying Volume and Pressure}
		\label{tab_moneyness_volume_NBP}
		\begin{tabular}{ccccccccccc}
			\toprule
			& \multicolumn{2}{c}{2017} & \multicolumn{2}{c}{2018} & \multicolumn{2}{c}{2019} & \multicolumn{2}{c}{2020} & \multicolumn{2}{c}{2021} \\
			& Call & Put & Call & Put & Call & Put & Call & Put & Call & Put \\
			\toprule
			& \multicolumn{10}{c}{Panel A. Net Buying Volume} \\
			DOTM & 35 & -380 & -1,646 & -1,035 & -21,544 & -20,385 & -64,920 & -15,895 & -233,394 & -164,373 \\
			OTM & -356 & 42 & -3,370 & -691 & -6,814 & -1,310 & -8,856 & -31,318 & 169,622 & -141,027 \\
			ATM & 959 & -209 & 469 & 290 & 8,691 & 3,757 & 36,455 & -13,410 & 207,075 & 233,671 \\
			ITM & 891 & 86 & 307 & -282 & 4,495 & 2,516 & 17,006 & 13,934 & 26,321 & -11,499 \\
			DITM & 391 & 25 & 351 & 895 & 945 & 487 & 12,689 & 1,459 & 18,442 & -2,174 \\
			Totals & 1,919 & -437 & -3,890 & -823 & -14,228 & -14,937 & -7,627 & -45,231 & 188,067 & -85,401 \\
			\toprule
			& \multicolumn{10}{c}{Panel B. Net Buying Pressure} \\
			DOTM & 6 & -21 & -150 & -61 & -1,340 & -1,256 & -3,777 & -781 & -18,221 & -14,559 \\
			OTM & -124 & 49 & -635 & -162 & -1,256 & -67 & -1,029 & -8,669 & 64,381 & -18,015 \\
			ATM & 455 & -84 & 234 & 123 & 4,270 & 2,016 & 19,432 & -3,575 & 94,818 & 108,546 \\
			ITM & 659 & 62 & 229 & -211 & 3,127 & 1,854 & 12,395 & 10,213 & 16,297 & -10,498 \\
			DITM & 364 & 22 & 316 & 837 & 875 & 453 & 11,745 & 1,361 & 17,614 & -2,430 \\
			Totals & 1,360 & 26 & -7 & 526 & 5,676 & 2,999 & 38,767 & -1,451 & 174,889 & 63,045 \\
			\toprule
			& \multicolumn{10}{c}{Panel C. Direction Net Buying Pressure for Calls} \\
			DOTM & \multicolumn{2}{c}{14} & \multicolumn{2}{c}{-44} & \multicolumn{2}{c}{-42} & \multicolumn{2}{c}{-1,498} & \multicolumn{2}{c}{-1,831} \\
			OTM & \multicolumn{2}{c}{-87} & \multicolumn{2}{c}{-236} & \multicolumn{2}{c}{-594} & \multicolumn{2}{c}{3,820} & \multicolumn{2}{c}{41,198} \\
			ATM & \multicolumn{2}{c}{270} & \multicolumn{2}{c}{55} & \multicolumn{2}{c}{1,127} & \multicolumn{2}{c}{11,504} & \multicolumn{2}{c}{-6,864} \\
			ITM & \multicolumn{2}{c}{299} & \multicolumn{2}{c}{220} & \multicolumn{2}{c}{637} & \multicolumn{2}{c}{1,091} & \multicolumn{2}{c}{13,397} \\
			DITM & \multicolumn{2}{c}{171} & \multicolumn{2}{c}{-261} & \multicolumn{2}{c}{211} & \multicolumn{2}{c}{5,192} & \multicolumn{2}{c}{10,022} \\
			Totals & \multicolumn{2}{c}{667} & \multicolumn{2}{c}{-266} & \multicolumn{2}{c}{1,338} & \multicolumn{2}{c}{20,109} & \multicolumn{2}{c}{55,922} \\
			\toprule
			& \multicolumn{10}{c}{Panel D.  Volatility Net Buying Pressure} \\
			DOTM & \multicolumn{2}{c}{-8} & \multicolumn{2}{c}{-105} & \multicolumn{2}{c}{-1,298} & \multicolumn{2}{c}{-2,279} & \multicolumn{2}{c}{-16,390} \\
			OTM & \multicolumn{2}{c}{-38} & \multicolumn{2}{c}{-398} & \multicolumn{2}{c}{-662} & \multicolumn{2}{c}{-4,849} & \multicolumn{2}{c}{23,183} \\
			ATM & \multicolumn{2}{c}{185} & \multicolumn{2}{c}{178} & \multicolumn{2}{c}{3,143} & \multicolumn{2}{c}{7,929} & \multicolumn{2}{c}{101,682} \\
			ITM & \multicolumn{2}{c}{360} & \multicolumn{2}{c}{9} & \multicolumn{2}{c}{2,490} & \multicolumn{2}{c}{11,304} & \multicolumn{2}{c}{2,900} \\
			DITM & \multicolumn{2}{c}{193} & \multicolumn{2}{c}{576} & \multicolumn{2}{c}{664} & \multicolumn{2}{c}{6,553} & \multicolumn{2}{c}{7,592} \\
			Totals & \multicolumn{2}{c}{693} & \multicolumn{2}{c}{260} & \multicolumn{2}{c}{4,337} & \multicolumn{2}{c}{18,658} & \multicolumn{2}{c}{118,967} \\
			\toprule
		\end{tabular}
		\begin{tablenotes}
			\item \small
			Note.   Average net buying volume and  net buying pressure in USD. Panels A and B represent the average of  hourly frequency net buying volume and net buying pressure     over an entire year.
		\end{tablenotes}
	\end{table}

	%
	%
	
	{\cite{gemmill1996did}, \cite{kelly2016too} and others demonstrate that the S\&P 500 options market is dominated by large, sophisticated institutional traders that demand puts to hedge   downside and crash risk. In line with this observation, \cite{bollen2004does} find that changes in implied volatility of S\&P 500 options are most 	strongly affected by buying pressure for put options. However, the interest of institutional traders in bitcoin options market is still at a very early stage. There are relatively more speculative traders, and these  have risk-aversion characteristics that may be quite different from the institutional traders in S\&P 500 options. That is our  main reason for finding support for volatility-motivated demand where none is present in S\&P 500 options. Furthermore, \cite{derman1999regimes} shows that the S\&P index has range-bounded, stable-trending and crash regimes. By contrast, bitcoin prices are prone to both upward and downward jumps. This is why we found that both volatility and directional motivated trades drive the bitcoin implied volatility curve for  calls as well as puts.}

	\section{Methodology}\label{sec_methods}
	\noindent To examine our net buying pressure hypotheses  we first test the limits-to-arbitrage versus learning hypotheses using the regression-based tests of \cite{bollen2004does}, also used by  \cite{kang2008information}. We then employ the test of \cite{chen2017net} to distinguish between volatility-motivated and directional-motivated demands. 	
	
	First we fix notation. Let $j\in \{C,P\}$ and denote the change in ATM call (resp. put) implied volatility from time $t-1$ to time $t$ by $\Delta \sigma _{j,t}^{\mbox{\tiny ATM}}$. Here, as in all moneyness categories, we obtain  a single value for $\sigma _{C,t}^{\mbox{\tiny ATM}}$ by averaging the implied volatilities over all ATM calls over each time interval $[t-1, t]$, and similarly for  $\sigma _{P,t}^{\mbox{\tiny ATM}}$. Now for $ k\in \left\{ \mbox{DOTM, OTM} \right\}$, let $\Delta \sigma _{j,t}^{k} $ denote the change from time $t-1$ to time $t$ in the average implied volatility of all call (resp. put) options in moneyness category $k$.   Let $r_t$ denote the return of the Deribit bitcoin spot index and $v_t$ denote the aggregated spot market volume from the exchanges in this index, both from time $t-1$ to $t$. We proxy this with the aggregated USD volumes from the three largest spot markets in the index, i.e. Bitstamp, Coinbase and Gemini. With this notation and the buying pressure variables $A_{j,t}^{\mbox{\tiny ATM}}$  defined in \eqref{eq_nbp_calculation}, the first two regressions of \cite{bollen2004does} may be written:
	\begin{align}\label{eq_bollen_atm} 
		\Delta \sigma _{j,t}^{\mbox{\tiny ATM}} &= \alpha_{0j} + \alpha_{1j} r_t +  \alpha_{2j} v_t +  \alpha_{3j} A_{C,t}^{\mbox{\tiny ATM}} + \alpha_{4j} A_{P,t}^{\mbox{\tiny ATM}} +  \alpha_{5j} \Delta \sigma _{j,{t-1}}^{\mbox{\tiny ATM}}+ \varepsilon_t,\quad j \in \{C,P\}. 
	\end{align}
	Including the contemporaneous underlying return controls for leverage effects, the trading volume controls for information flow effects, { and the lagged implied volatility captures mean-reversion so that significance of $\alpha_{5C}$ and/or $\alpha_{5P}$ supports the limits-to-arbitrage hypothesis.}
	
	If we cannot {reject the hypothesis  $\alpha_{3j} = \alpha_{4j}$, for either $j=C$ or $j=P$, then trading is strongly motivated by expectations about volatility; this is because volatility traders have no reason  to prefer ATM calls to ATM puts or conversely. However, when $\alpha_{3C}$ and $\alpha_{4P}$ are both significant and positive this  does not necessarily imply that there is no volatility learning, even if we do reject that $\alpha_{3j} = \alpha_{4j}$ for either calls or puts. It only means that the idiosyncratic demand for ATM calls is different from that for ATM puts.}  If we do find evidence in favour of a learning hypothesis then, following  \cite{kang2008information} we can distinguish which of volatility or directional learning is the dominant hypothesis by running six further regressions of the form:
	
	\begin{align}\label{eq_bollen_all_moneyness} 
		\Delta \sigma _{j,t}^{k}  &= \alpha_{0j}^k + \alpha_{1j}^k r_t + \alpha_{2j}^k v_t + \alpha_{3j}^k A_{j,t}^{k} +\alpha_{4j}^k A_{i,t}^{\mbox{\tiny ATM}} + \alpha_{5j}^k \Delta \sigma _{j,{t-1}}^{k}+ \varepsilon_t^k, \quad i, j \in \{C,P\} . 
	\end{align}
	for $k\in \left\{ \mbox{DOTM, OTM} \right\}$.
	{The regressors  {again} include the lagged change in implied volatility and its sign and significance  distinguishes  limits-to-arbitrage from the dominant learning hypothesis, i.e. that market makers learn from the prevalent type of trading, being volatility motivated or directional motivated.  In this case, changes in implied volatility should be permanent and
		uncorrelated through time, hence that $\alpha_{5j}=0$ should not be rejected. A significant estimate of $\alpha_{4j}^k$ for both $k = \mbox{OTM}$ and $k=\mbox{DOTM}$ indicates that demand for ATM options affects  whole level of the implied volatility curve.}   Under the directional-learning
	hypothesis informed traders would act  oppositely    in calls and puts,  so we expect significant values for  $\alpha_{3C}^k$ and/or $\alpha_{4P}^k$ with opposite signs, regardless of moneyness $k$. The 	volatility-learning hypothesis predicts that 
	$\alpha_{4j}^k> \alpha_{3j}^k> 0 $ for calls and/or puts, because ATM options have the greatest vega so traders should  react most rapidly to ATM buying pressures. We remark that  it is also possible  to have $\alpha_{3j}^k> \alpha_{4j}^k> 0 $  under the volatility-learning hypothesis, but this would only happen if the idiosyncratic volatility-motivated demand for $k$-moneyness options  dominate the volatility-motivated demand on ATM options.

	The methodology proposed by \cite{kang2008information} can only assess which is the dominating learning hypothesis, but the model proposed by \cite{chen2017net} and later employed by \cite{ryu2021impact} allows one to  distinguish between the two types of learning hypotheses by running regressions of the form: 
	\begin{align}\label{eq_reg_chan}
		\Delta \sigma _{j,t}^{k}&=\beta_{0j}^k +\beta_{1j}^k r_t+\beta_{2j}^k v_t +\beta_{3j}^k V_{t}^{k}+\beta_{4j}^k  D_{j,t}^{k}+\beta _{5j}^k\Delta \sigma_{j,t-1}^{k}+\eta_{j,t}^k,
	\end{align}
	where $ j\in \{Call, Put\}=\{C,P\}$ and $k\in \{ATM, OTM, DOTM\}$. The parameters are similar to \eqref{eq_bollen_all_moneyness}, except that we replace net buying pressure on ATM options by two different testable components, i.e.  directional-motivated demand \eqref{eq_direction_NBP} and volatility-motivated demand \eqref{eq_vol_NBP}. The size and significance of $\beta_{3j}^k$ and $\beta_{4j}^k$ reveals which of the two learning hypotheses predominates and the sign of $\beta_{4j}^k$ also indicates the direction of the demand pressure on $k$-moneyness options of type $j$.

	\section{Empirical Results}\label{sec_empirical}
	As Table \ref{tab_moneyness_volume_byyear} shows, the trading volume on Deribit bitcoin options has increased tremendously over time: from only 0.26 USD billion  in   2017  to 102.96 USD billion  in January to July of 2021. {This  rapid increase  in trading volume supports our previous finding that  the implied volatility curves for bitcoin options have been converging to a shape very similar to that of the  S\&P 500 implied volatility curves -- see Figure \ref{moneyness_compare}.  This indicates that bitcoin   options prices since 2019  are likely to be more reliable and not distorted by stale prices and low trading  volumes.} Our empirical results for 2017 and 2018 provide weak support for the limits-to-arbitrage hypothesis but do not show any evidence of either directional or volatility learning. For brevity, we do not present them here, but instead we focus on results  for 2019, for 2020 and for the first 7 months of 2021  -- recalling that data were incomplete after July 2021, since they were confined to short-term options only.
	
	We use implied volatility and net buying pressure measures at the hourly frequency but our results are robust to several other data frequencies, such as 4 or  8 hours but not if we aggregate to the daily frequency. {We aggregate to hourly data because Deribit options are not as actively traded as major equity index options, and also because a daily window is too wide to detect informed trading in such a fast-changing and volatile market.\footnote{{The daily results are available on request. Our main findings are that the limits-to-arbitrage hypothesis is mainly supported by put options; that volatility learning is more pronounced in 2019 and 2020; and that only directional learning is supported in 2021 –- which is understandable due to the escalating of bitcoin price during the 2021  sample. }} Specifically,  \cite{bouoiyour2016drives}, \cite{goczek2019drives}  and \cite{brauneis2021measure}  all show that liquidity can be adversely affected by exogenous shocks such as large `whale' bitcoin movements on blockchains, changes in investors sentiment recorded on social media, or regulatory policy and economy policy uncertainties. The effect of such shocks can sometimes be very short lived, because of the extreme volatility of bitcoin. During the course of one day much information flows into the market and the inventory managed by the market maker also changes considerably, so we do not expect to identify the same buying pressure effects on implied volatility at the daily frequency.}
	
	\begin{table}[h!]
		\centering
		\caption{The   Test  of  \cite{bollen2004does} for ATM, OTM, and DOTM Options}
		\label{tab_bollen_test_others}
		\begin{threeparttable}\small 
			\begin{tabular}{ccccccccccc}
				\toprule
				& Year &    & $ \alpha _0 $ & $ \alpha _1 $ & $ \alpha _2 $ & $ \alpha _3 $ & $ \alpha _4 $ & $ \alpha _5 $ & Nobs \\
				\toprule 
				&   &       & \multicolumn{6}{c}{Panel A. ATM Options}     &     \\
				\multirow{3}{*}{$\Delta \sigma^{\mbox{\tiny ATM}}_{C}$} & 2019 &       & -0.27$^{***}$ & 6.609 & 0.067$^{***}$ & 3.072$^{***}$ & 1.264 & -0.458$^{***}$ & 5868 \\
				& 2020    &  & -0.58$^{***}$ & -0.58$^{***}$ & 0.116$^{***}$ & 0.403$^{**}$ & 0.317$^{*}$ & -0.426$^{***}$ & 8315 \\
				& 2021   &  & -0.677$^{***}$ & -31.985$^{***}$ & 0.041$^{***}$ & 0.2$^{***}$ & 0.015 & -0.395$^{***}$ & 4192 \\ [1em]
				\multirow{3}{*}{$\Delta \sigma^{\mbox{\tiny ATM}}_{P}$} & 2019 &     & -0.348$^{***}$ & -21.161$^{**}$ & 0.073$^{***}$ & 0.015 & 7.594$^{***}$ & -0.413$^{***}$ & 5034 \\
				& 2020 &    & -0.608$^{***}$ & -50.548$^{***}$ & 0.128$^{***}$ & 0.326$^{*}$ & 0.674$^{***}$ & -0.399$^{***}$ & 7883 \\
				& 2021 &   & -0.635$^{***}$ & -46.709$^{***}$ & 0.038$^{***}$ & 0.127$^{***}$ & -0.116$^{*}$ & -0.397$^{***}$ & 4094 \\	
				\toprule
				&   &  $i$ \mbox{in} (\ref{eq_bollen_all_moneyness})       &  \multicolumn{6}{c}{Panel B. OTM Options}     &     \\
				\multirow{6}{*}{$\Delta \sigma^{\mbox{\tiny OTM}}_{C}$} & \multirow{2}{*}{2019}   & $Call $ & -0.202$^{***}$ & -16.439$^{*}$ & 0.07$^{***}$ & 4.188$^{***}$ & 0.925 & -0.443$^{***}$ & 6548 \\
				&  &     $Put $ & -0.194$^{***}$ & -12.071 & 0.068$^{***}$ & 4.25$^{***}$ & 1.5$^{*}$ & -0.443$^{***}$ & 6548 \\
				& \multirow{2}{*}{2020}     & $Call $ & -0.569$^{***}$ & -13.064 & 0.122$^{***}$ & 1.19$^{***}$ & 0.284 & -0.431$^{***}$ & 8581 \\
				&  &    $Put $ & -0.563$^{***}$ & -8.446 & 0.122$^{***}$ & 1.298$^{***}$ & 0.421$^{**}$ & -0.431$^{***}$ & 8581 \\
				& \multirow{2}{*}{2021}     & $Call $ & -0.717$^{***}$ & -28.898$^{***}$ & 0.042$^{***}$ & 0.364$^{***}$ & 0.179$^{***}$ & -0.386$^{***}$ & 4343 \\
				&     & $Put $ & -0.721$^{***}$ & -25.284$^{***}$ & 0.043$^{***}$ & 0.406$^{***}$ & 0.058 & -0.386$^{***}$ & 4343 \\
				&    &  &  &  &  &  &  &  &  \\
				\multirow{6}{*}{$\Delta \sigma^{\mbox{\tiny OTM}}_{P}$} & \multirow{2}{*}{2019}  & $Call $ & -0.3$^{***}$ & 6.613 & 0.079$^{***}$ & 6.706$^{***}$ & 0.351 & -0.444$^{***}$ & 6049 \\
				&  &  $Put $   & -0.292$^{***}$ & 9.174 & 0.076$^{***}$ & 6.552$^{***}$ & 1.342 & -0.444$^{***}$ & 6050 \\
				& \multirow{2}{*}{2020}    & $Call $ & -0.487$^{***}$ & -72.922$^{***}$ & 0.105$^{***}$ & 0.972$^{***}$ & 0.385$^{**}$ & -0.406$^{***}$ & 8446 \\
				&  &     $Put $ & -0.487$^{***}$ & -68.261$^{***}$ & 0.107$^{***}$ & 0.819$^{***}$ & 0.346$^{*}$ & -0.406$^{***}$ & 8446 \\
				& \multirow{2}{*}{2021} &   $Call $ & -0.551$^{***}$ & -28.553$^{***}$ & 0.034$^{***}$ & 0.255$^{***}$ & 0.143$^{***}$ & -0.387$^{***}$ & 4293 \\
				&  &     $Put $ & -0.544$^{***}$ & -26.629$^{***}$ & 0.035$^{***}$ & 0.241$^{***}$ & -0.058 & -0.388$^{***}$ & 4293 \\
				\toprule
				&   &       &  \multicolumn{6}{c}{Panel C.  DOTM Options}     &     \\
				\multirow{6}{*}{$\Delta \sigma^{\mbox{\tiny DOTM}}_{C}$} & \multirow{2}{*}{2019}  & $Call $ & -0.412$^{***}$ & -26.902$^{**}$ & 0.109$^{***}$ & 23.144$^{***}$ & -0.76 & -0.43$^{***}$ & 5040 \\
				&       & $Put $ & -0.404$^{***}$ & -25.886$^{**}$ & 0.103$^{***}$ & 22.97$^{***}$ & 1.762 & -0.43$^{***}$ & 5041 \\
				& \multirow{2}{*}{2020}     & $Call $ & -0.517$^{***}$ & -12.616 & 0.12$^{***}$ & 6.851$^{***}$ & 0.288 & -0.466$^{***}$ & 7748 \\
				&    & $Put $ & -0.51$^{***}$ & -5.841 & 0.12$^{***}$ & 7.131$^{***}$ & 0.608$^{**}$ & -0.466$^{***}$ & 7748 \\
				& \multirow{2}{*}{2021}     & $Call $ & -0.737$^{***}$ & -23.409$^{**}$ & 0.047$^{***}$ & 2.098$^{***}$ & 0.13$^{**}$ & -0.419$^{***}$ & 4214 \\
				&      & $Put $ & -0.729$^{***}$ & -21.621$^{**}$ & 0.048$^{***}$ & 2.114$^{***}$ & -0.072 & -0.42$^{***}$ & 4214 \\
				&    &  &  &  &  &  &  &  &  \\
				\multirow{6}{*}{$\Delta \sigma^{\mbox{\tiny DOTM}}_{P}$} & \multirow{2}{*}{2019}   & $Call $ & -0.243$^{*}$ & 19.27 & 0.085$^{***}$ & 28.457$^{***}$ & -0.161 & -0.422$^{***}$ & 4420 \\
				&  &    $Put $ & -0.223 & 23.283 & 0.077$^{***}$ & 28.134$^{***}$ & 3.139$^{**}$ & -0.423$^{***}$ & 4421 \\
				& \multirow{2}{*}{2020}    & $Call $ & -0.503$^{***}$ & -107.839$^{***}$ & 0.105$^{***}$ & 5.056$^{***}$ & 0.368 & -0.42$^{***}$ & 7849 \\
				&     & $Put $ & -0.505$^{***}$ & -103.141$^{***}$ & 0.107$^{***}$ & 4.926$^{***}$ & 0.123 & -0.42$^{***}$ & 7849 \\
				& \multirow{2}{*}{2021}    & $Call $ & -0.513$^{***}$ & -43.96$^{***}$ & 0.029$^{***}$ & 1.613$^{***}$ & 0.123$^{*}$ & -0.406$^{***}$ & 4223 \\
				&      & $Put $ & -0.513$^{***}$ & -41.181$^{***}$ & 0.03$^{***}$ & 1.592$^{***}$ & 0.01 & -0.406$^{***}$ & 4223 \\
				\toprule
			\end{tabular}
			
			\begin{tablenotes}
				\item \small 
				Note.  {The top panel conducts regressions in \eqref{eq_bollen_atm} for ATM call and put options from 2019 to 2021 and  should be read as $\Delta \sigma _{call, t}^{\mbox{\tiny ATM}} = \alpha_{0} + \alpha_{1} r_t +  \alpha_{2} v_t +  \alpha_{3} A_{call,t}^{\mbox{\tiny ATM}} + \alpha_{4} A_{put,t}^{\mbox{\tiny ATM}} +  \alpha_{5} \Delta \sigma _{call,{t-1}}^{\mbox{\tiny ATM}}+ \varepsilon_t.$ The middle and bottom panels report regressions in \eqref{eq_bollen_all_moneyness} for   OTM and DOTM     options and are interpreted similarly.  The limits-to-arbitrage hypothesis is consistently supported by the highly significant coefficient $\alpha_5$.  If directional learning is supported, we would observe $\alpha_3>0$ and $\alpha_4 <0$ since traders would act oppositely on calls and puts whereas volatility learning is more pronounced if both $\alpha_3$ and $\alpha_4$ are positive.}  
			$^{*}, ^{**}, ^{***}$ indicates 10\%, 5\% and 1\% significance levels respectively.   
		\end{tablenotes}
	\end{threeparttable}
\end{table}

To differentiate limits-to-arbitrage from volatility/directional learning  hypotheses, we first run the \cite{bollen2004does} test for ATM, OTM, and DOTM    option categories, examining the sign and magnitude of parameters $\alpha_3, \alpha_4$ and $\alpha_5$. Table \ref{tab_bollen_test_others} reports the results. Several interesting facts are observed. { First,  in line with the findings of \cite{bollen2004does}, \cite{kang2008information} and \cite{chen2017net} for well-developed option markets, the limits-to-arbitrage  hypothesis is consistently supported across moneyness categories. The lagged implied volatility regression coefficient $\alpha_5$ estimates are negative and   highly significant with a decaying trend. For instance,  in the ATM call options category, the coefficient $\alpha_5$   decreases  from $-0.458$  in 2019 to $-0.395$  in  2021 that is roughly double the size of the corresponding coefficient estimates found for well-developed  index options markets. For instance, a value of around -0.17 is reported for the S\&P 500 options market in \cite{bollen2004does} and  \cite{chen2017net} find a value of about -0.23 for TAIEX index options.  This shows that bitcoin options market makers manage inventory less effectively than market makers for S\&P 500 and TAIEX options. However, as the bitcoin options market matures, we could expect trading costs to diminish along with the heterogeneity of skills and knowledge between bitcoin traders. As this maturing process continues in bitcoin markets we expect the the magnitude of the limits-to-arbitrage coefficient $\alpha_5$ to decrease continually.}

Secondly, the estimate of {the informational effect} $\alpha_2$ is positive and significant at 1\%, for all moneyness categories in every year. Thus, spot market trading volume has a positive effect on the change in implied volatility, which  is not a surprising as trading volume in bitcoin spot markets is much larger than it is on Deribit options. Nevertheless, the controlled leverage effect ($\alpha_1$) generally becomes more significant over time, {so Deribit is gradually attracting more participants to take highly leveraged positions. 
	Thirdly, for ATM options, the coefficients $\alpha_3$  and $\alpha_4$   are both positive, for calls and for puts, except for a weakly significant   $\alpha_{4P}<0$ in 2021 which is probably   due to the surge in bitcoin prices which rose from around 30,000 USD in January 2021 to above 63,000 USD in April 2021, then crashed back below 30,000 USD in July 2021. 
	Similarly, except for a weakly significant   $\alpha_{4P}<0$ in 2021, we always have $\alpha_3>\alpha_4>0$ for OTM and DOTM options.   Hence, the estimate of  idiosyncratic net buy pressure $\alpha_3$   consistently drives the changes in implied volatility levels of these options, and dominates the force of ATM net buying pressure.  This shows that although ATM options have the highest volatility sensitivity (vega)  they  do not effectively drag the whole implied volatility curve along OTM and DOTM categories.   Taken together, we can infer that  volatility learning  dominates directional learning.  Additionally,  the gradually decreasing magnitudes of  $\alpha_3$  and $\alpha_4$ year by year indicates a maturing of the  market where  both option prices and IV levels  are more impacted by informed trading.}


\begin{table}[h!]
	\centering
	\caption{The   Test  of  \cite{chen2017net} for ATM, OTM, and DOTM  Options}
	\label{tab_chen_wang_test_others}
	\begin{threeparttable}\small 
		\begin{tabular}{cccccccccc}
			\toprule
			&   &  $ \beta _0 $ & $ \beta_1 $ & $ \beta_2 $ & $ \beta _3 $ & $ \beta _4 $ & $ \beta_5 $ & Nobs \\
			\toprule
			&  &  & \multicolumn{6}{c}{Panel A. ATM Options} &  \\
			\multirow{3}{*}{$\Delta \sigma_C^{\mbox{\tiny ATM}}$} & 2019  & -0.27$^{***}$ & 6.609 & 0.067$^{***}$ & 4.336$^{***}$ & 1.808 & -0.458$^{***}$ & 5868 \\
			& 2020    & -0.58$^{***}$ & -13.841 & 0.116$^{***}$ & 0.72$^{***}$ & 0.086 & -0.426$^{***}$ & 8315 \\
			& 2021    & -0.677$^{***}$ & -31.985$^{***}$ & 0.041$^{***}$ & 0.215$^{***}$ & 0.184$^{**}$ & -0.395$^{***}$ & 4192 \\
			[1em]
			\multirow{3}{*}{$\Delta \sigma_P^{\mbox{\tiny ATM}}$} & 2019   & -0.348$^{***}$ & -21.161$^{**}$ & 0.073$^{***}$ & 7.609$^{***}$ & 7.58 & -0.413$^{***}$ & 5034 \\
			& 2020    & -0.608$^{***}$ & -50.548$^{***}$ & 0.128$^{***}$ & 1.0$^{***}$ & 0.349 & -0.399$^{***}$ & 7883 \\
			& 2021    & -0.635$^{***}$ & -46.709$^{***}$ & 0.038$^{***}$ & 0.011 & -0.243$^{***}$ & -0.397$^{***}$ & 4094 \\
			\toprule
			&  &  & \multicolumn{6}{c}{Panel B. OTM Options} &  \\
			\multirow{3}{*}{$\Delta \sigma_C^{\mbox{\tiny OTM}}$} & 2019  & -0.188$^{**}$ & -9.277 & 0.067$^{***}$ & 6.594$^{***}$ & 1.685 & -0.443$^{***}$ & 6550 \\
			& 2020   & -0.564$^{***}$ & -8.791 & 0.122$^{***}$ & 1.599$^{***}$ & 0.982$^{***}$ & -0.431$^{***}$ & 8581 \\
			& 2021    & -0.715$^{***}$ & -24.309$^{***}$ & 0.043$^{***}$ & 0.484$^{***}$ & 0.335$^{***}$ & -0.386$^{***}$ & 4343 \\
			[1em]
			\multirow{3}{*}{$\Delta \sigma_P^{\mbox{\tiny OTM}}$} & 2019  & -0.299$^{***}$ & 6.379 & 0.079$^{***}$ & 7.135$^{***}$ & 6.22$^{***}$ & -0.444$^{***}$ & 6051 \\
			& 2020    & -0.482$^{***}$ & -73.318$^{***}$ & 0.106$^{***}$ & 1.439$^{***}$ & 0.512$^{**}$ & -0.407$^{***}$ & 8446 \\
			& 2021    & -0.551$^{***}$ & -30.256$^{***}$ & 0.034$^{***}$ & 0.426$^{***}$ & 0.063 & -0.388$^{***}$ & 4293 \\
			\toprule
			&   &  & \multicolumn{6}{c}{Panel C. DOTM Options} &  \\
			\multirow{3}{*}{$\Delta \sigma_C^{\mbox{\tiny DOTM}}$} & 2019  & -0.408$^{***}$ & -22.74$^{*}$ & 0.109$^{***}$ & 33.172$^{***}$ & 11.951$^{*}$ & -0.43$^{***}$ & 5042 \\
			& 2020    & -0.519$^{***}$ & -10.118 & 0.122$^{***}$ & 6.97$^{***}$ & 7.018$^{***}$ & -0.466$^{***}$ & 7747 \\
			& 2021    & -0.733$^{***}$ & -18.559$^{*}$ & 0.048$^{***}$ & 2.493$^{***}$ & 1.756$^{***}$ & -0.42$^{***}$ & 4214 \\
			[1em]
			\multirow{3}{*}{$\Delta \sigma_P^{\mbox{\tiny DOTM}}$} & 2019   & -0.23 & 11.777 & 0.089$^{***}$ & 43.519$^{***}$ & 11.671 & -0.424$^{***}$ & 4422 \\
			& 2020    & -0.496$^{***}$ & -105.627$^{***}$ & 0.106$^{***}$ & 6.521$^{***}$ & 3.878$^{**}$ & -0.42$^{***}$ & 7848 \\
			& 2021    & -0.505$^{***}$ & -42.78$^{***}$ & 0.03$^{***}$ & 1.958$^{***}$ & 1.249$^{**}$ & -0.407$^{***}$ & 4221 \\
			\toprule
		\end{tabular}
		\begin{tablenotes}
			\item \small 
			Note.  This table reports regressions     in \eqref{eq_reg_chan} for ATM, OTM, and DOTM    options.  {For instance, in ATM options category, each row should be read as $$\Delta \sigma _{call,t}^{\mbox{\tiny ATM}}=\beta_{0}^{\mbox{\tiny ATM}} +\beta_{1}^{\mbox{\tiny ATM}}  r_t+\beta_{2}^{\mbox{\tiny ATM}} v_t +\beta_{3}^{\mbox{\tiny ATM}} V_{t}^{\mbox{\tiny ATM}}+\beta_{4}^{\mbox{\tiny ATM}}  D_{t}^{\mbox{\tiny ATM}}+\beta _{5}^{\mbox{\tiny ATM}}\Delta \sigma_{call, t-1}^{\mbox{\tiny ATM}}+\eta_{t}^{\mbox{\tiny ATM}}.$$ The $^{*}, ^{**}, ^{***}$ indicate 10\%, 5\% and 1\% significance levels respectively.  Any significance of coefficients $\beta_3$ and $\beta_4$ demonstrates the existence of volatility and directional learning respectively. }
			
		\end{tablenotes}
	\end{threeparttable}
\end{table}

To  investigate the possible co-existence of  volatility and directional learning  we run the \cite{chen2017net} tests  in \eqref{eq_reg_chan} for ATM, OTM, and DOTM   options by turning off the channel of net-buying-pressure from ATM options. The results are displayed in Table \ref{tab_chen_wang_test_others}. Since for ATM, OTM and DOTM options the estimate of $\beta_3$ is always greater than that of $\beta_4$,  volatility-motivated trades consistently dominate directional-motivated trades.  This finding is consistent with the results in Table \ref{tab_bollen_test_others} but contrasts findings of previous studies on well-developed markets, where directional learning is generally more pronounced than volatility learning, see \cite{bollen2004does}, \cite{kang2008information}, \cite{chen2017net} and \cite{ryu2021impact}. {Additionally, the sign, significance,
	and magnitude of  $\beta_4$ for directional trading varies across different  moneyness categories, probably because the bitcoin options market is still
	maturing. Similar findings about net-buying-pressure varying for
	different option categories are presented by \cite{kang2008information} for  KOSPI 200 index options
	and by \cite{chen2017net} for TAIEX options.} 

{Although the magnitudes of volatility and directional net buying pressures differ, we could treat  the ratio of   $\beta_3/\beta_4$ in regressions (\ref{eq_reg_chan})      as the relative strength of the two driving forces.  For example, for the
	TAIEX options market, \cite{chen2017net}  find that the ratio   $\beta_3/\beta_4$   for OTM options is close to $2:5$,
	which implies that the volatility trading effect is approximately 40\% of the directional trading
	effect for OTM options (see also the discussion of Table VII on Page 18 in \cite{chen2017net}). Our results   paint a different picture for bitcoin
	options markets. We estimate a ratio 
	$\beta_3/\beta_4$   about 3:2 for OTM calls, which shows that the volatility learning effect is roughly
	1.5 times larger than the directional learning effect. So the evidence supporting the learning
	hypothesis in bitcoin, is very different from the evidence from established stock index options markets.
	That is, much more of the informed trading in bitcoin options is about volatility rather than
	direction. Our economic intuition behind this is that it is likely caused by, or could even
	cause, the high volatility of bitcoin prices. Unfortunately, the regressions we perform are not
	designed to detect the direction of causality, and that is beyond the scope of this paper, but
	it would be   an interesting subject for further research.}

However, it should be mentioned that only directional learning is evidenced by ATM puts in  2021, and this is likely due to the bitcoin price bubble as already noted above.  
In 2019 and 2020, both volatility and directional-motivated demands  drive OTM and DOTM implied volatility changes, while only volatility   information drives ATM option trades. The intuition here is that informed traders with privileged  volatility   and directional information would   prefer trading  cheaper OTM and DOTM options instead of  more expensive ATM options.  This finding is in line with those for the KOSPI 200 options market \citep{ryu2021impact} and the  TAIEX index options market  \citep{chen2017net}. But it contrasts the  S\&P 500 index options results of \cite{bollen2004does}, the HSI index options results of \cite{chan2004net} and the KOSPI 200 index options results of \cite{kang2008information}. 
Finally, the limits-to-arbitrage hypothesis is consistently supported, as previously demonstrated by the  regressions \eqref{eq_bollen_atm} and \eqref{eq_bollen_all_moneyness}: the estimates of information flow and leverage controls ($\beta_1$ and $\beta_2$) and of $\beta_5$  are all similar to the values estimated for $\alpha_1, \alpha_2$ and $\alpha_5$ in \cite{bollen2004does} test of Table \ref{tab_bollen_test_others}.

\begin{table}[t]
	\centering
	\caption{The   \cite{chen2017net} Test for Options with Different  Maturities:  2019}
	\label{tab_chen_wang_ttm_2019}
	\begin{threeparttable}\small 
		\begin{tabular}{ccccccccc}
			\toprule
			Model &  Maturity & $ \beta _0 $ & $ \beta_1  $ & $ \beta_2  $ & $ \beta _3  $ & $ \beta _4  $ & $ \beta_5 $ & Nobs \\
			\toprule
			&  & \multicolumn{6}{c}{Panel A. ATM Options} &  \\
			\multirow{3}{*}{$\Delta \sigma_C^{\mbox{\tiny ATM}}$} & $[1,7]$ & -0.596$^{***}$ & -7.087 & 0.136$^{***}$ & 6.855$^{***}$ & 5.322$^{**}$ & -0.345$^{***}$ & 2662 \\
			& $[8,21]$ & -0.397$^{***}$ & 13.453$^{*}$ & 0.097$^{***}$ & 7.989$^{***}$ & -3.077 & -0.279$^{***}$ & 2056 \\
			& $\geq 22$ & -0.076 & -28.118$^{***}$ & 0.029$^{**}$ & 1.857 & 2.12 & -0.474$^{***}$ & 2453 \\
			[1em]
			\multirow{3}{*}{$\Delta \sigma_P^{\mbox{\tiny ATM}}$} & $[1,7]$ & -0.73$^{***}$ & 8.035 & 0.192$^{***}$ & 1.004 & 6.376$^{*}$ & -0.374$^{***}$ & 2140 \\
			& $[8,21]$ & -0.549$^{***}$ & -65.33$^{***}$ & 0.137$^{***}$ & 9.182$^{***}$ & 1.25 & -0.391$^{***}$ & 1688 \\
			& $\geq 22$ & -0.258$^{**}$ & -23.146$^{**}$ & 0.041$^{***}$ & 6.627$^{***}$ & 6.648$^{***}$ & -0.371$^{***}$ & 1297 \\
			\toprule
			&  & \multicolumn{6}{c}{Panel B. OTM Options} &  \\
			\multirow{3}{*}{$\Delta \sigma_C^{\mbox{\tiny OTM}}$} & $[1,7]$ & -0.811$^{***}$ & -48.143$^{***}$ & 0.203$^{***}$ & 22.084$^{***}$ & 3.808 & -0.32$^{***}$ & 3253 \\
			& $[8,21]$ & -0.356$^{***}$ & -15.165$^{**}$ & 0.09$^{***}$ & 8.202$^{***}$ & 3.634$^{*}$ & -0.352$^{***}$ & 2509 \\
			& $\geq 22$ & -0.07 & -3.898 & 0.04$^{***}$ & 7.561$^{***}$ & 3.148 & -0.469$^{***}$ & 2974 \\
			[1em]
			\multirow{3}{*}{$\Delta \sigma_P^{\mbox{\tiny OTM}}$} & $[1,7]$ & -0.379$^{**}$ & 20.606 & 0.092$^{***}$ & 11.193$^{***}$ & 13.645$^{***}$ & -0.226$^{***}$ & 2765 \\
			& $[8,21]$ & -0.365$^{***}$ & 11.133 & 0.085$^{***}$ & 8.962$^{***}$ & 5.624$^{**}$ & -0.372$^{***}$ & 2170 \\
			& $\geq 22$ & -0.199$^{**}$ & -3.165 & 0.043$^{***}$ & 7.887$^{***}$ & 4.44$^{**}$ & -0.525$^{***}$ & 2770 \\
			\toprule
			&  & \multicolumn{6}{c}{Panel C. DOTM Options} &  \\
			\multirow{3}{*}{$\Delta \sigma_C^{\mbox{\tiny DOTM}}$} & $[1,7]$ & -0.496$^{**}$ & 3.193 & 0.165$^{***}$ & 95.855$^{***}$ & 24.032 & -0.345$^{***}$ & 1942 \\
			& $[8,21]$ & -0.475$^{***}$ & -36.932$^{***}$ & 0.118$^{***}$ & 12.059 & 43.791$^{***}$ & -0.371$^{***}$ & 1291 \\
			& $\geq 22$ & -0.155 & -1.535 & 0.048$^{***}$ & 26.81$^{***}$ & 16.425$^{*}$ & -0.418$^{***}$ & 1907 \\
			[1em]
			\multirow{3}{*}{$\Delta \sigma_P^{\mbox{\tiny DOTM}}$} & $[1,7]$ & -0.476$^{*}$ & 74.258$^{***}$ & 0.156$^{***}$ & 103.254$^{***}$ & 135.639$^{***}$ & -0.267$^{***}$ & 1585 \\
			& $[8,21]$ & -0.412$^{**}$ & 5.732 & 0.106$^{***}$ & 52.84$^{***}$ & 18.197 & -0.483$^{***}$ & 1011 \\
			& $\geq 22$ & -0.219 & 18.968 & 0.061$^{***}$ & 10.825 & 26.332$^{***}$ & -0.41$^{***}$ & 1367
			\\
			\toprule
		\end{tabular}
		
		\begin{tablenotes}
			\item \small 
			Note.  This table reports regressions in  \eqref{eq_reg_chan}  for     subsets of   options by maturity, using the terms `short-term', `medium-term',  and `longer-term'  for  options with maturities lying in $[1,7]$, $[8,21]$ and $\geq 22$ days.
			The $^{*}, ^{**}, ^{***}$ indicate 10\%, 5\% and 1\% significance levels respectively. 
		\end{tablenotes}
	\end{threeparttable}
\end{table}

\begin{table}[h!]
	\centering
	\caption{The     \cite{chen2017net} Test for Options with Different  Maturities:  2020}
	\label{tab_chen_wang_ttm_2020}
	\begin{threeparttable}\small 
		\begin{tabular}{ccccccccc}
			\toprule
			Model &  Maturity & $ \beta _0 $ & $ \beta_1  $ & $ \beta_2  $ & $ \beta _3  $ & $ \beta _4  $ & $ \beta_5 $ & Nobs \\
			\toprule
			&  & \multicolumn{6}{c}{Panel A. ATM Options} &  \\
			\multirow{3}{*}{$\Delta \sigma_C^{\mbox{\tiny ATM}}$} & $[1,7]$ & -1.03$^{***}$ & -28.32$^{***}$ & 0.201$^{***}$ & 1.717$^{***}$ & 0.779 & -0.314$^{***}$ & 6834 \\
			& $[8,21]$ & -0.45$^{***}$ & -6.295 & 0.081$^{***}$ & 1.264$^{***}$ & 0.215 & -0.267$^{***}$ & 4507 \\
			& $\geq 22$ & -0.173$^{**}$ & -1.605 & 0.032$^{***}$ & 0.338 & 0.009 & -0.434$^{***}$ & 4756 \\
			[1em]
			\multirow{3}{*}{$\Delta \sigma_P^{\mbox{\tiny ATM}}$} & $[1,7]$ & -0.89$^{***}$ & -96.113$^{***}$ & 0.182$^{***}$ & 1.98$^{***}$ & 1.046$^{*}$ & -0.332$^{***}$ & 6148 \\
			& $[8,21]$ & -0.43$^{***}$ & -103.303$^{***}$ & 0.091$^{***}$ & 1.916$^{***}$ & 0.56 & -0.165$^{***}$ & 3386 \\
			& $\geq 22$ & -0.281$^{**}$ & -9.382 & 0.054$^{***}$ & 0.598 & 0.23 & -0.353$^{***}$ & 2909 \\
			\toprule
			&  & \multicolumn{6}{c}{Panel B. OTM Options} &  \\
			\multirow{3}{*}{$\Delta \sigma_C^{\mbox{\tiny OTM}}$} & $[1,7]$ & -0.873$^{***}$ & -37.728$^{***}$ & 0.187$^{***}$ & 5.168$^{***}$ & 2.942$^{***}$ & -0.315$^{***}$ & 7408 \\
			& $[8,21]$ & -0.445$^{***}$ & -25.821$^{***}$ & 0.084$^{***}$ & 1.801$^{***}$ & 1.504$^{***}$ & -0.32$^{***}$ & 4862 \\
			& $\geq 22$ & -0.215** & -40.337$^{***}$ & 0.041$^{***}$ & 0.623 & 0.74$^{*}$ & -0.437$^{***}$ & 5346 \\
			[1em]
			\multirow{3}{*}{$\Delta \sigma_P^{\mbox{\tiny OTM}}$} & $[1,7]$ & -0.905$^{***}$ & -122.33$^{***}$ & 0.194$^{***}$ & 4.979$^{***}$ & 2.022$^{***}$ & -0.314$^{***}$ & 7089 \\
			& $[8,21]$ & -0.401$^{***}$ & -68.098$^{***}$ & 0.079$^{***}$ & 2.226$^{***}$ & 0.822$^{**}$ & -0.306$^{***}$ & 4739 \\
			& $\geq 22$ & -0.211$^{**}$ & -33.249$^{***}$ & 0.041$^{***}$ & 0.589 & 0.122 & -0.444$^{***}$ & 5031 \\
			\toprule
			&  & \multicolumn{6}{c}{Panel C. DOTM Options} &  \\
			\multirow{3}{*}{$\Delta \sigma_C^{\mbox{\tiny DOTM}}$} & $[1,7]$ & -1.0$^{***}$ & -70.308$^{***}$ & 0.196$^{***}$ & 24.28$^{***}$ & 17.258$^{***}$ & -0.423$^{***}$ & 5452 \\
			& $[8,21]$ & -0.465$^{***}$ & -59.095$^{***}$ & 0.084$^{***}$ & 9.26$^{***}$ & 7.233$^{***}$ & -0.347$^{***}$ & 2839 \\
			& $\geq 22$ & -0.113 & -35.514$^{***}$ & 0.051$^{***}$ & 1.733 & 5.558$^{***}$ & -0.449$^{***}$ & 3895 \\
			[1em]
			\multirow{3}{*}{$\Delta \sigma_P^{\mbox{\tiny DOTM}}$} & $[1,7]$ & -0.705$^{***}$ & -135.8$^{***}$ & 0.16$^{***}$ & 20.181$^{***}$ & 14.017$^{***}$ & -0.389$^{***}$ & 5704 \\
			& $[8,21]$ & -0.452$^{***}$ & -70.164$^{***}$ & 0.08$^{***}$ & 11.789$^{***}$ & 6.511$^{**}$ & -0.407$^{***}$ & 3132 \\
			& $\geq 22$ & -0.163 & -50.448$^{***}$ & 0.036$^{***}$ & 3.667$^{*}$ & 2.756 & -0.464$^{***}$ & 3465
			\\
			\toprule
		\end{tabular}
		
		\begin{tablenotes}
			\item \small 
			Note.  This table reports regressions in  \eqref{eq_reg_chan}  for     subsets of   options by maturity, using the terms `short-term', `medium-term',  and `longer-term'  for  options with maturities lying in $[1,7]$, $[8,21]$ and $\geq 22$ days.
			The $^{*}, ^{**}, ^{***}$ indicate 10\%, 5\% and 1\% significance levels respectively.   
		\end{tablenotes}
	\end{threeparttable}
\end{table}

\begin{table}[h!]
	\centering
	\caption{The     \cite{chen2017net} Test for Options with Different  Maturities:  2021}
	\label{tab_chen_wang_ttm_2021}
	\begin{threeparttable}\small 
		\begin{tabular}{ccccccccc}
			\toprule
			Model &  Maturity & $ \beta _0 $ & $ \beta_1 $ & $ \beta_2   $ & $ \beta _3  $ & $ \beta _4  $ & $ \beta_5 $ & Nobs \\
			\toprule
			&  & \multicolumn{6}{c}{Panel A. ATM Options} &  \\
			\multirow{3}{*}{$\Delta \sigma_C^{\mbox{\tiny ATM}}$} & $[1,7]$ & -1.206$^{***}$ & -78.355$^{***}$ & 0.075$^{***}$ & 0.153 & 0.442$^{**}$ & -0.316$^{***}$ & 3620 \\
			& $[8,21]$ & -0.563$^{***}$ & -10.161$^{*}$ & 0.032$^{***}$ & 0.291$^{***}$ & 0.439$^{***}$ & -0.22$^{***}$ & 2898 \\
			& $\geq 22$ & -0.218$^{***}$ & 6.894 & 0.013$^{***}$ & 0.121 & 0.286$^{***}$ & -0.406$^{***}$ & 3332 \\
			[1em]
			\multirow{3}{*}{$\Delta \sigma_P^{\mbox{\tiny ATM}}$} & $[1,7]$ & -0.936$^{***}$ & -70.334$^{***}$ & 0.059$^{***}$ & -0.222 & -0.823$^{***}$ & -0.295$^{***}$ & 3324 \\
			& $[8,21]$ & -0.711$^{***}$ & -37.339$^{***}$ & 0.04$^{***}$ & 0.319$^{***}$ & -0.098 & -0.172$^{***}$ & 2506 \\
			& $\geq 22$ & -0.358$^{**}$ & -14.85 & 0.02$^{***}$ & -0.256 & -0.506$^{***}$ & -0.434$^{***}$ & 2552 \\
			\toprule
			&  & \multicolumn{6}{c}{Panel B. OTM Options} &  \\
			\multirow{3}{*}{$\Delta \sigma_C^{\mbox{\tiny OTM}}$} & $[1,7]$ & -1.113$^{***}$ & -71.348$^{***}$ & 0.072$^{***}$ & 0.9$^{***}$ & 0.695$^{***}$ & -0.273$^{***}$ & 3996 \\
			& $[8,21]$ & -0.654$^{***}$ & -7.892 & 0.036$^{***}$ & 0.66$^{***}$ & 0.345$^{***}$ & -0.265$^{***}$ & 3400 \\
			& $\geq 22$ & -0.245$^{***}$ & -5.492 & 0.015$^{***}$ & 0.301$^{**}$ & 0.3$^{**}$ & -0.402$^{***}$ & 3592 \\
			[1em]
			\multirow{3}{*}{$\Delta \sigma_P^{\mbox{\tiny OTM}}$} & $[1,7]$ & -0.995$^{***}$ & -115.481$^{***}$ & 0.068$^{***}$ & 1.145$^{***}$ & 0.663$^{***}$ & -0.288$^{***}$ & 3911 \\
			& $[8,21]$ & -0.552$^{***}$ & -45.935$^{***}$ & 0.03$^{***}$ & 0.468$^{***}$ & -0.044 & -0.344$^{***}$ & 3317 \\
			& $\geq 22$ & -0.323$^{***}$ & 0.47 & 0.017$^{***}$ & 0.287 & 0.016 & -0.465$^{***}$ & 3565 \\
			\toprule
			&  & \multicolumn{6}{c}{Panel C. DOTM Options} &  \\
			\multirow{3}{*}{$\Delta \sigma_C^{\mbox{\tiny DOTM}}$} & $[1,7]$ & -1.293$^{***}$ & -52.329$^{***}$ & 0.078$^{***}$ & 4.582$^{***}$ & 5.204$^{***}$ & -0.399$^{***}$ & 3417 \\
			& $[8,21]$ & -0.714$^{***}$ & -5.671 & 0.041$^{***}$ & 3.454$^{***}$ & 2.808$^{***}$ & -0.352$^{***}$ & 2563 \\
			& $\geq 22$ & -0.316$^{***}$ & -8.414 & 0.02$^{***}$ & 2.698$^{***}$ & 1.212$^{*}$ & -0.473$^{***}$ & 2922 \\
			[1em]
			\multirow{3}{*}{$\Delta \sigma_P^{\mbox{\tiny DOTM}}$} & $[1,7]$ & -0.809$^{***}$ & -139.774$^{***}$ & 0.061$^{***}$ & 5.505$^{***}$ & 2.618$^{**}$ & -0.372$^{***}$ & 3438 \\
			& $[8,21]$ & -0.57$^{***}$ & -38.043$^{***}$ & 0.032$^{***}$ & 1.437$^{*}$& 3.195$^{***}$ & -0.376$^{***}$ & 2634 \\
			& $\geq 22$ & -0.315 & -4.755 & 0.01 & 2.882$^{**}$ & 1.477 & -0.459$^{***}$ & 2800
			\\
			\toprule
		\end{tabular}
		
		\begin{tablenotes}
			\item \small 
			Note.  This table reports regressions in  \eqref{eq_reg_chan}     for     subsets of   options by maturity, using the terms `short-term', `medium-term',  and `longer-term'  for  options with maturities lying in $[1,7]$, $[8,21]$ and $\geq 22$ days.
			The $^{*}, ^{**}, ^{***}$ indicate 10\%, 5\% and 1\% significance levels respectively. 
		\end{tablenotes}
	\end{threeparttable}
\end{table}

Next we apply the methodology of \cite{chen2017net} to  subsets of   options by maturity, using the terms `short-term', `medium-term',  and `longer-term'  for  options with maturities lying in $[1,7]$, $[8,21]$ and $\geq 22$ days. 
In general, we find that the volatility-informed traders  prefer short-term and medium-term  options.   In 2020,  trading on longer-term ATM options  only supports the limits-to-arbitrage hypothesis, whereas it is directional trading that dominates prices of the longer-term OTM calls, and volatility information that drives the prices of longer-term DOTM puts. Although more directional trading emerges in  longer-term options during 2021, less informed trading is present in longer-term options overall, than in short and medium-term  options.

{The tick-level data also allows investigation of time-of-day effects, for which there are conflicting results from previous research. For instance, \cite{jain2019insights}  found bitcoin spot trading volume is higher on centralised exchanges during their local working hours  and that there was substantial weekend trading, both reflecting high retail participation. However,  \cite{alexander2021volatility}   found that most volatility transmission between bitcoin perpetual swaps and spot markets occurs during US and Asian trading times. Both these papers focused on spot and futures only. Our next results are the first to investigate time-of-day effects of buying pressures in bitcoin options markets.}

\begin{table}[h!]
	\centering \small
	\caption{Net Buying Volume and Net Buying Pressure by Time of Day}
	\label{tab_timeofday_NBP}
	\begin{center}
		\begin{tabular}{ccccccc}
			\toprule
			Time of   Day (UTC) & \multicolumn{2}{c}{00:00 -- 08:00} & \multicolumn{2}{c}{08:00 -- 16:00} & \multicolumn{2}{c}{16:00 -- 24:00} \\
			Year & Call & Put & Call & Put & Call & Put \\
			\toprule
			& \multicolumn{6}{c}{Panel A. Net Buying Pressure} \\
			2017 & 8.45 & 0.05 & 13.87 & -1.15 & 13.66 & 1.69 \\
			2018 & -14.98 & -8.89 & 6.68 & 9.33 & 13.52 & 7.02 \\
			2019 & 40.03 & 35.53 & 60.42 & 18.20 & 55.49 & 3.56 \\
			2020 & 331.96 & -134.49 & 804.47 & 473.04 & 118.93 & -358.11 \\
			2021 & 1854.06 & 459.91 & 1856.20 & 1070.60 & 448.06 & 42.02 \\
			\toprule
			& \multicolumn{6}{c}{Panel B. Direction Net Buying Pressure for Calls} \\
			2017 & \multicolumn{2}{c}{4.20} & \multicolumn{2}{c}{7.51} & \multicolumn{2}{c}{5.98} \\
			2018 & \multicolumn{2}{c}{-3.05} & \multicolumn{2}{c}{-1.32} & \multicolumn{2}{c}{3.25} \\
			2019 & \multicolumn{2}{c}{2.25} & \multicolumn{2}{c}{21.11} & \multicolumn{2}{c}{25.96} \\
			2020 & \multicolumn{2}{c}{233.23} & \multicolumn{2}{c}{165.72} & \multicolumn{2}{c}{238.52} \\
			2021 & \multicolumn{2}{c}{697.08} & \multicolumn{2}{c}{396.61} & \multicolumn{2}{c}{203.02} \\
			\toprule
			& \multicolumn{6}{c}{Panel C. Volatility Net Buying Pressure} \\
			2017 & \multicolumn{2}{c}{4.25} & \multicolumn{2}{c}{6.36} & \multicolumn{2}{c}{7.68} \\
			2018 & \multicolumn{2}{c}{-11.93} & \multicolumn{2}{c}{8.00} & \multicolumn{2}{c}{10.27} \\
			2019 & \multicolumn{2}{c}{37.78} & \multicolumn{2}{c}{39.31} & \multicolumn{2}{c}{29.53} \\
			2020 & \multicolumn{2}{c}{98.73} & \multicolumn{2}{c}{638.75} & \multicolumn{2}{c}{-119.59} \\
			2021 & \multicolumn{2}{c}{1156.99} & \multicolumn{2}{c}{1467.21} & \multicolumn{2}{c}{245.04}\\
			\toprule
		\end{tabular}
	\end{center}
	\begin{tablenotes}
		\item \small
		Note. Panel A: Average of the 8-hour net buying pressure measures for calls and puts  over different years in three time zones: 00:00 -- 08:00 (Asia), 08:00 -- 16:00 (Europe) and 16:00 -- 00:00 (US). Panels B and C disaggregate of the net buying pressure shown in Panel A into directional-motivated and volatility-motivated net buying pressure. For calls, the entry in Panel A is the sum of the corresponding entries in Panels B and C. For puts, the entry in Panel A is the difference between the corresponding entries in Panels C and B. 
	\end{tablenotes}
\end{table}

Table \ref{tab_timeofday_NBP} reports the net buying pressure, and the directional- and volatility-motivated buying pressures by the time-of-day, where we group results into  8-hour time slots and take the average during each year. The time slots of 00:00 -- 08:00, 08:00 -- 16:00, and 16:00 -- 24:00 UTC are chosen because they roughly correspond to Asian, European and US trading hours, respectively. It is really noticeable that option trader's net buying pressure  is much lower during US trading times. Also distinct time-of-day pattern emerges in 2020 and 2021, where directional-motivated demand is greatest during Asian trading times and volatility-motivated demand most evident during European trading times. 	{But further tests, reported in Table \ref{tab_chen_wang_test_time_of_day}, show that volatility-motivated demand,   mainly driven by ATM options, is greatest during US and Asian trading times.  This finding is also supported by the volatility-transmission results of  \cite{alexander2021volatility}. So although Table \ref{tab_timeofday_NBP} shows that the greatest volatility-motivated demand occurs during European trading times, the information content of this demand is relatively low. Another take-away from the OTM and DOTM results in Table \ref{tab_chen_wang_test_time_of_day} is that directional learning is only supported  during Asian   and US trading times.}

\begin{table}[h!]\small
	\centering
	\caption{ {The Time-of-Day Effect of    Volatility and Directional Learning}}
	\label{tab_chen_wang_test_time_of_day}
	\begin{tabular}{cccccccc}
		\toprule
		& UTC Time   & $ \beta _0 $ & $ \beta_1 $ & $ \beta_2 $ & $ \beta _3 $ & $ \beta _4 $ & $ \beta_5 $ \\
		\toprule
		&   & \multicolumn{6}{c}{Panel A. ATM Options} \\
		\multirow{3}{*}{$\Delta   \sigma_C^{\tiny \mbox{ATM}}$} & $(0,8]$  & -4.367$^{***}$ & -9.196 & 0.057$^{***}$ & 0.417$^{*}$ & 0.222 & -0.094$^{*}$ \\
		& $(8,16]$ &    0.636 & 1.099 & 0.03$^{***}$ & 0.327 & -0.561 & -0.17$^{***}$ \\
		& $(16,24]$ &    -1.301$^{***}$ & -53.767$^{***}$ & 0.044$^{***}$ & 0.74$^{**}$ & 0.461 & -0.112$^{**}$ \\
		[1em]
		\multirow{3}{*}{$\Delta   \sigma_P^{\tiny \mbox{ATM}}$} & $(0,8]$   & -4.937$^{***}$ & -74.652$^{***}$ & 0.066$^{***}$ & 0.268 & -0.005 & -0.001 \\
		& $(8,16]$  & 0.682 & 7.111 & 0.039$^{***}$ & 0.329 & 0.581 & -0.191$^{***}$ \\
		& $(16,24]$    & -1.907$^{***}$ & -163.991$^{***}$ & 0.081$^{***}$ & 1.177$^{***}$ & -0.28 & -0.255$^{***}$ \\
		\toprule
		&   & \multicolumn{6}{c}{Panel B. OTM Options} \\
		\multirow{3}{*}{$\Delta   \sigma_C^{\tiny \mbox{OTM}}$} & $(0,8]$   & -3.469$^{***}$ & -4.27 & 0.049$^{***}$ & 0.76$^{***}$ & 0.959$^{***}$ & -0.069 \\
		& $(8,16]$ &    1.186$^{***}$ & 25.891$^{*}$ & 0.03$^{***}$ & 0.001 & 0.242 & 0.002 \\
		& $(16,24]$ &   -2.278$^{***}$ & -79.752$^{***}$ & 0.069$^{***}$ & 2.044$^{***}$ & 1.697$^{***}$ & -0.137$^{**}$ \\
		[1em]
		\multirow{3}{*}{$\Delta   \sigma_P^{\tiny \mbox{OTM}}$} & $(0,8]$   & -3.535$^{***}$ & -85.011$^{***}$ & 0.051$^{***}$ & 0.886$^{***}$ & -0.231 & 0.055 \\
		& $(8,16]$ &   0.471 & -57.316$^{***}$ & 0.043$^{***}$ & 0.264 & 0.123 & -0.159$^{***}$ \\
		& $(16,24]$ &   -2.165$^{***}$ & -131.82$^{***}$ & 0.077$^{***}$ & 1.852$^{***}$ & -0.85$^{***}$ & -0.144$^{***}$ \\
		\toprule
		&    & \multicolumn{6}{c}{Panel C. DOTM Options} \\
		\multirow{3}{*}{$\Delta   \sigma_C^{\tiny \mbox{DOTM}}$} & $(0,8]$   & -3.891$^{***}$ & -4.825 & 0.048$^{***}$ & 3.802$^{*}$ & 6.51$^{***}$ & -0.294$^{***}$ \\
		& $(8,16]$ &   0.415 & 18.39 & 0.041$^{***}$ & 6.021$^{***}$ & 0.596 & -0.024 \\
		& $(16,24]$ &   -1.287$^{**}$ & -92.179$^{***}$ & 0.083$^{***}$ & 8.909$^{***}$ & 9.339$^{***}$ & -0.257$^{***}$ \\
		[1em]
		\multirow{3}{*}{$\Delta   \sigma_P^{\tiny \mbox{DOTM}}$} & $(0,8]$  & -4.127$^{***}$ & -188.93$^{***}$ & 0.05$^{***}$ & 5.759$^{***}$ & -2.15 & -0.281$^{***}$ \\
		& $(8,16]$ &    0.328 & -88.713$^{***}$ & 0.035$^{***}$ & 5.088$^{***}$ & -0.381 & -0.112$^{***}$ \\
		& $(16,24]$ &    -1.033 & -309.699$^{***}$ & 0.104$^{***}$ & 5.362 & -6.475$^{*}$& -0.368$^{***}$\\
		\toprule
	\end{tabular}
	\begin{tablenotes}
		\item \small 
		Note.  {Table \ref{tab_timeofday_NBP} shows the
			time-of-day pattern only really emerged after 2020. Therefore, in this table we run the regressions in (\ref{eq_reg_chan})
			using data from 2020 and 2021 only.  } 
		The $^{*}, ^{**}, ^{***}$ indicate 10\%, 5\% and 1\% significance levels respectively.   
	\end{tablenotes}
\end{table}

\section{Conclusion}\label{sec_conclusion}
{Compared with the well-developed S\&P 500 options market which is also severely regulated, we found numerous differences in the bitcoin options market which is dominated by the `self-regulated' exchange, Deribit. S\&P 500 options implied volatility curves have exhibited a marked `skew' for decades, and the ATM volatility has been around 20\% most of the time. By contrast, the bitcoin implied volatility curve is much higher, with ATM volatility often around 100\% but the level is also very variable, and the curve is much flatter and more symmetric than the S\&P 500 `skew' shape. All these differences point to quite different data generation processes for the underlying prices. For instance, the features we have documented in bitcoin implied volatilities indicate that the stochastic volatility models driving bitcoin prices should include  positive as well as negative jumps, which are correlated with the variance process.}

{However, the focus of our study in this paper is not on option pricing but on the information we can derive from trading patterns in bitcoin options, by modelling their implied volatilities. The existence of informed trading is clearly supported by our regressions of returns, and realised volatility and implied volatility on their own lag and the lagged trading volume of Deribit bitcoin options.} We examine the limits-to-arbitrage hypothesis for market maker's supply, and the volatility-  and directional-motivated demand-side effects on hourly changes in the implied volatility of Deribit bitcoin options, using different types of net-buying-pressure metrics. By  contrast with previous results on well-developed stock and index options markets, our results show that not only limits-to-arbitrage constraints but also volatility-informed demand are the main driving forces behind ATM options, {although  the net buying pressure from ATM options does not appear to drag the whole implied volatility curve along when  ATM volatility changes.} Deribit trading volumes have increased enormously during the last few years and market makers are using these options to manage order imbalance and inventory more effectively as the  market matures. Also most traders are motivated to pursue volatility information for the purpose of hedging, arbitraging and speculating.   

The bitcoin market is highly volatile and no  consensus on the fundamentals   of bitcoin prices is widely acknowledged, so it is extremely difficult for most traders to exploit advanced information about the direction of bitcoin price movement. We find that directional learning, as well as  volatility learning, is  supported  by our results for OTM and DOTM options which are much less expensive than ATM options. Directional learning effects are more pronounced during 2021, even being present in   ATM put options. We attribute this to the price bubble and crash which may have been precipitated by the behaviour of a few large highly-informed traders.

Further refinements of our tests assess  time-to-maturity and time-of-day effects. We find that volatility-informed trading dominates demand pressures  on short-term and medium-term  options.     Although more directional trading is apparent in  longer-term options during 2021, overall  they encompass   less informed trades than short- and medium-term  options.  {The time-of-day patterns  indicate that net buying pressure grew rapidly during 2020 and 2021, that directional pressure is strongest during Asian trading times and volatility pressure is strongest during European times. As in traditional option markets we find strong evidence of market makers managing inventory under limits-to-arbitrage. However, in contrast with traditional markets, we also find that bitcoin market makers adjust prices using information from both directional and volatility motivated trades, especially by observing demand pressures during Asian and US trading times.}\\

\bigskip
\bigskip
\singlespacing


\newpage

\bibliographystyle{apalike}
\bibliography{reference}

\begin{thebibliography}{}

\bibitem[Alexander et~al., 2020]{alexander2020bitmex}
Alexander, C., Choi, J., Park, H., and Sohn, S. (2020).
\newblock Bitmex bitcoin derivatives: Price discovery, informational
  efficiency, and hedging effectiveness.
\newblock {\em Journal of Futures Markets}, 40(1):23--43.

\bibitem[Alexander et~al., 2021a]{alexander2021optimal}
Alexander, C., Deng, J., and Zou, B. (2021a).
\newblock Hedging with bitcoin futures: The effect of liquidation loss aversion
  and aggressive trading.
\newblock {\em arXiv preprint arXiv:2101.01261}.

\bibitem[Alexander et~al., 2021b]{alexander2021volatility}
Alexander, C., Heck, D., and Kaeck, A. (2021b).
\newblock The role of binance in bitcoin volatility transmission.
\newblock {\em arXiv preprint arXiv:2107.00298}.

\bibitem[Bakshi et~al., 2003]{bakshi2003stock}
Bakshi, G., Kapadia, N., and Madan, D. (2003).
\newblock Stock return characteristics, skew laws, and the differential pricing
  of individual equity options.
\newblock {\em The Review of Financial Studies}, 16(1):101--143.

\bibitem[Bollen and Whaley, 2004]{bollen2004does}
Bollen, N.~P. and Whaley, R.~E. (2004).
\newblock Does net buying pressure affect the shape of implied volatility
  functions?
\newblock {\em The Journal of Finance}, 59(2):711--753.

\bibitem[Bouoiyour et~al., 2016]{bouoiyour2016drives}
Bouoiyour, J., Selmi, R., Tiwari, A.~K., Olayeni, O.~R., et~al. (2016).
\newblock What drives bitcoin price.
\newblock {\em Economics Bulletin}, 36(2):843--850.

\bibitem[Brauneis et~al., 2021]{brauneis2021measure}
Brauneis, A., Mestel, R., Riordan, R., and Theissen, E. (2021).
\newblock How to measure the liquidity of cryptocurrency markets?
\newblock {\em Journal of Banking \& Finance}, 124:106041.

\bibitem[Chan et~al., 2004]{chan2004net}
Chan, K.~C., Cheng, L.~T., and Lung, P.~P. (2004).
\newblock Net buying pressure, volatility smile, and abnormal profit of hang
  seng index options.
\newblock {\em Journal of Futures Markets}, 24(12):1165--1194.

\bibitem[Cheah and Fry, 2015]{cheah2015speculative}
Cheah, E.-T. and Fry, J. (2015).
\newblock Speculative bubbles in bitcoin markets? an empirical investigation
  into the fundamental value of bitcoin.
\newblock {\em Economics Letters}, 130:32--36.

\bibitem[Chen and Wang, 2017]{chen2017net}
Chen, C.-C. and Wang, S.-H. (2017).
\newblock Net buying pressure and option informed trading.
\newblock {\em Journal of Futures Markets}, 37(3):238--259.

\bibitem[Chuang et~al., 2020]{chuang2020impact}
Chuang, Y.-W., Tsai, W.-C., and Wu, M.-H. (2020).
\newblock The impact of net buying pressure on vix option prices.
\newblock {\em Journal of Futures Markets}, 40(2):209--227.

\bibitem[Cong et~al., 2020]{cong2020crypto}
Cong, L.~W., Li, X., Tang, K., and Yang, Y. (2020).
\newblock Crypto wash trading.
\newblock {\em Available at SSRN 3530220}.

\bibitem[Demir et~al., 2018]{demir2018does}
Demir, E., Gozgor, G., Lau, C. K.~M., and Vigne, S.~A. (2018).
\newblock Does economic policy uncertainty predict the bitcoin returns? an
  empirical investigation.
\newblock {\em Finance Research Letters}, 26:145--149.

\bibitem[Deng et~al., 2020]{deng2020minimum}
Deng, J., Pan, H., Zhang, S., and Zou, B. (2020).
\newblock Minimum-variance hedging of bitcoin inverse futures.
\newblock {\em Applied Economics}, 52(58):6320--6337.

\bibitem[Derman, 1999]{derman1999regimes}
Derman, E. (1999).
\newblock Regimes of volatility.
\newblock {\em Risk}, pages 55--59.

\bibitem[Easley et~al., 1998]{easley1998option}
Easley, D., O'hara, M., and Srinivas, P.~S. (1998).
\newblock Option volume and stock prices: Evidence on where informed traders
  trade.
\newblock {\em The Journal of Finance}, 53(2):431--465.

\bibitem[Gemmill, 1996]{gemmill1996did}
Gemmill, G. (1996).
\newblock Did option traders anticipate the crash? evidence from volatility
  smiles in the uk with us comparisons.
\newblock {\em The Journal of Futures Markets (1986-1998)}, 16(8):881.

\bibitem[Goczek and Skliarov, 2019]{goczek2019drives}
Goczek, {\L}. and Skliarov, I. (2019).
\newblock What drives the bitcoin price? a factor augmented error correction
  mechanism investigation.
\newblock {\em Applied Economics}, 51(59):6393--6410.

\bibitem[Griffin and Shams, 2020]{griffin2020bitcoin}
Griffin, J.~M. and Shams, A. (2020).
\newblock Is bitcoin really untethered?
\newblock {\em The Journal of Finance}, 75(4):1913--1964.

\bibitem[Jain et~al., 2019]{jain2019insights}
Jain, P.~K., McInish, T.~H., and Miller, J.~L. (2019).
\newblock Insights from bitcoin trading.
\newblock {\em Financial Management}, 48(4):1031--1048.

\bibitem[Kang and Park, 2008]{kang2008information}
Kang, J. and Park, H.-J. (2008).
\newblock The information content of net buying pressure: Evidence from the
  kospi 200 index option market.
\newblock {\em Journal of Financial Markets}, 11(1):36--56.

\bibitem[Kelly et~al., 2016]{kelly2016too}
Kelly, B., Lustig, H., and Van~Nieuwerburgh, S. (2016).
\newblock Too-systemic-to-fail: What option markets imply about sector-wide
  government guarantees.
\newblock {\em American Economic Review}, 106(6):1278--1319.

\bibitem[Lakonishok et~al., 2007]{lakonishok2007option}
Lakonishok, J., Lee, I., Pearson, N.~D., and Poteshman, A.~M. (2007).
\newblock Option market activity.
\newblock {\em The Review of Financial Studies}, 20(3):813--857.

\bibitem[Makarov and Schoar, 2020]{makarov2020trading}
Makarov, I. and Schoar, A. (2020).
\newblock Trading and arbitrage in cryptocurrency markets.
\newblock {\em Journal of Financial Economics}, 135(2):293--319.

\bibitem[Ryu et~al., 2021]{ryu2021impact}
Ryu, D., Ryu, D., and Yang, H. (2021).
\newblock The impact of net buying pressure on index options prices.
\newblock {\em Journal of Futures Markets}, 41(1):27--45.

\bibitem[Scaillet et~al., 2020]{scaillet2020high}
Scaillet, O., Treccani, A., and Trevisan, C. (2020).
\newblock High-frequency jump analysis of the bitcoin market.
\newblock {\em Journal of Financial Econometrics}, 18(2):209--232.

\bibitem[Shen et~al., 2019]{shen2019does}
Shen, D., Urquhart, A., and Wang, P. (2019).
\newblock Does twitter predict bitcoin?
\newblock {\em Economics Letters}, 174:118--122.

\bibitem[Shiu et~al., 2010]{shiu2010impact}
Shiu, Y.-M., Pan, G.-G., Lin, S.-H., and Wu, T.-C. (2010).
\newblock Impact of net buying pressure on changes in implied volatility:
  Before and after the onset of the subprime crisis.
\newblock {\em The Journal of Derivatives}, 17(4):54--66.

\bibitem[Siu and Elliott, 2021]{siu2021bitcoin}
Siu, T.~K. and Elliott, R.~J. (2021).
\newblock Bitcoin option pricing with a setar-garch model.
\newblock {\em The European Journal of Finance}, 27(6):564--595.

\bibitem[Tiwari et~al., 2019]{tiwari2019modelling}
Tiwari, A.~K., Kumar, S., and Pathak, R. (2019).
\newblock Modelling the dynamics of bitcoin and litecoin: Garch versus
  stochastic volatility models.
\newblock {\em Applied Economics}, 51(37):4073--4082.

\bibitem[Yan, 2011]{yan2011jump}
Yan, S. (2011).
\newblock Jump risk, stock returns, and slope of implied volatility smile.
\newblock {\em Journal of Financial Economics}, 99(1):216--233.

\bibitem[Zhang et~al., 2021]{zhang2021downside}
Zhang, W., Li, Y., Xiong, X., and Wang, P. (2021).
\newblock Downside risk and the cross-section of cryptocurrency returns.
\newblock {\em Journal of Banking \& Finance}, 133:106246.

\end{thebibliography}


%
%
%
%
%
%
%
%
%
%
%
%
%

\end{document}